\documentclass[12pt]{article}
\usepackage{a4wide}
\usepackage{amssymb}

\newcommand{\ratn}{\mbox{\(\mathbb Q\)}}

\newcommand{\reals}{\mbox{\(\mathbb R\)}}

\newcommand{\unit}{[0,1]}

\newcommand{\clos}{\mbox{\rm Cl}}
\newcommand{\mos}{\mbox{\rm mos}}
\newcommand{\cat}{\;^\wedge}

\newcommand{\lead}{\mbox{\rm lead}}
\newcommand{\trail}{\mbox{\rm trail}}
\newcommand{\shuffle}{\mbox{\rm shuffle}}

\newcommand{\blank}{\#}

\newcommand{\start}{\mbox{\rm start}}
\newcommand{\cover}{\mbox{\rm cover}}
\newcommand{\fin}{\mbox{\rm end}}

\newcommand{\sqpp}{*_q^{\phi}(\phi)}

\newcommand{\limit}{\mbox{\rm limit}}
\newcommand{\tick}{\mbox{\rm tick}}

\newcommand{\truth}{\top}
\newcommand{\falsity}{\bot}

\newcommand{\lang}{L(U,S)}

\newcommand{\power}[1]{\wp( #1 )}

\newcommand{\SetFigFont}[5]{} 
\renewcommand{\smash}[1]{#1}

\newtheorem{lemma}{LEMMA}
\newtheorem{theorem}{THEOREM}
\newtheorem{definition}{DEFINITION} 

\newtheorem{claim}{CLAIM}

\newenvironment{proof}{
\begin{quotation}PROOF:}{
$\square$ \end{quotation}}

\newenvironment{nemma}[1]{
\begin{lemma}
\label{lem:#1}}{
\end{lemma}}

\newenvironment{naim}[1]{
\begin{claim}
\label{claim:#1}}{
\end{claim}}

\newcommand{\sfproof}{
}

\newcommand{\prorem}[1]{}

\title{
The Complexity of Temporal Logic
over the Reals}
\author{M. Reynolds\\
Murdoch University, Australia}

\begin{document}

\maketitle

\begin{abstract}
It is shown that the decision problem
for the temporal logic with  until
and since connectives
over real-numbers time
is PSPACE-complete.
\end{abstract}

\section{Introduction}

There are a variety of temporal logics
appropriate for a variety of 
reasoning tasks.
Propositional reasoning on a natural
numbers model of time has been 
well studied via the
logic now commonly called PLTL 
which was introduced in \cite{Pnu77}.
However, it has long been
acknowledged that dense or specifically
real-numbers time models
may be better for many applications,
ranging from philosophical, natural language
and AI modelling of human reasoning
to computing and engineering applications of 
concurrency, refinement, open systems,
analogue devices and metric information.
See for example
\cite{KMP94} or \cite{BuG85}.

The most natural and useful such temporal logic
is propositional 
temporal logic over real-numbers time
using the Until and Since
connectives introduced in \cite{Kam68}.
We will call this logic RTL in this paper.
We know from \cite{Kam68} that this
logic is sufficiently expressive for many applications:
technically it is expressively complete
and so at least as expressive
as any other usual temporal logic which could
be defined over real-numbers time
and as expressive as the first-order
monadic logic of the real numbers.
We have, from 
\cite{GaHo90} and \cite{Rey:R},
complete axiom systems to allow
derivation of the validities of RTL.
We know from \cite{BuG85} that RTL is
decidable, ie that an algorithm exists 
for deciding whether a given RTL
formula is a validity or not.
Unfortunately, it has seemed
difficult to develop the 
reasoning procedures any further.
It is not even clear from the
decision procedure in \cite{BuG85}
(via Rabin's non-elementarily complex
decision procedure for the second-order
monadic logic of two successors)
how computationally complex 
it might be to decide validities in RTL.

This is in marked contrast to the
situation with PLTL which has been
shown to have a  PSPACE-complete
decision problem in \cite{SiC85}.
A variety of practical reasoning methods for
PLTL have been developed.

Here we show that 
as far as determining 
validity
is concerned,
RTL is just as easy to reason
with as PLTL.
In particular, the complexity
of the decision problem
is PSPACE-complete.

This opens the way 
for the development
of efficient reasoning procedures for RTL
and
for many practical applications.
For example, it is commonly required
to determine 
consequence relations
between finite sets of
formulas,
eg a detailed description
of the running of
a system
and a desirable overall property.
Such a question is equivalent
to a validity question.

Note that there has been some work
on restricted versions of temporal logic over 
the reals. In
\cite{Rab98} and
\cite{KMP94} the assumption
of finite variability is made,
ie it is supposed that atoms do not change
their truth values densely in time.
Under such an assumption, standard
discrete time techniques can be used to
develop decision procedures. 
We do not make any such assumptions.

The proof here uses new techniques in
temporal logic.
In particular we further develop
the idea of linear time mosaics
as seen in \cite{Rey:cult}.
Mosaics were used to prove
decidability
of certain theories of relation algebras
in \cite{Nem95} and have been
used since quite generally in algebraic
logic and modal logic.
These mosaics are small pieces of a model,
in our case, a small piece of a real-flowed structure.
We decide whether a finite set of small pieces
is sufficient to be used to build
a real-numbers model of a given formula.
This is also equivalent to
the existence of a winning strategy for one player
in a two-player game played with mosaics.
The search for a winning strategy can be
arranged into a search through a tree
of mosaics which we can proceed through
in a depth-first manner.
By establishing limits on the depth
of the tree (a polynomial in terms
of the length of the formula)
and on the branching factor
(exponential) we can ensure
that we have a PSPACE algorithm
as we only need to remember a small
fixed amount of information about
all the previous siblings of a given node.

In the case of the real numbers
in this paper we do not emphasize the
game aspect of this search but 
instead study certain structures
which correspond to tactics in the game.
By ensuring that mosaics get
simpler as we get deeper in the tree
we can respect the depth bound
and also capture the Dedekind completeness
of the underlying flow.
By ensuring that certain thorough mutually
recursive relationships (called shuffles)
between mosaics in the trees include
at least one very simple pair of mosaics,
we can also capture the 
separability property of the reals.

The proof also vaguely suggests
a tableau based
method for determining
validity but developing such
a method will need some more work.

\section{The logic}
\label{sec:logic}

Fix a countable set $\cal L$
of atoms.
Here, frames $(T,<)$,
or flows of time, will
be irreflexive linear orders.
Structures ${\cal T}=(T,<,h)$ 
will have  a frame $(T,<)$
and a valuation $h$
for the atoms
i.e. for each atom $p \in {\cal L}$,
$h(p) \subseteq T$.
Of particular importance will be
{\em real} structures
${\cal T}=(\reals,<,h)$ which have the real numbers flow
(with their usual irreflexive linear ordering). 

The language $L(U,S)$  is generated by the
2-place connectives
$U$ and $S$
along with classical $\neg$ and $\wedge$.
That is, we define the set of formulas
recursively to contain
the atoms
%and $\truth$ (for truth)
and for formulas $\alpha$ and $\beta$ we
include $\neg \alpha$, $\alpha \wedge \beta$,
$U(\alpha,\beta)$ and $S(\alpha,\beta)$.

Formulas are evaluated at points in 
structures ${\cal T}=(T,<,h)$.
We write
${\cal T}, x \models \alpha$
when $\alpha$ is true at the point $x \in T$. 
This is defined
recursively as follows.
Suppose that  we have defined
the truth of formulas $\alpha$ and $\beta$
at all points of
$\cal T$. Then for all points $x$:\\
\begin{tabular}{lll}
${\cal T},  x \models p$ & iff & $x \in h(p)$, for $p$
atomic;\\ 
%${\cal T},  x \models \truth$; &  &\\
${\cal T},  x \models \neg \alpha$ & iff &
${\cal T},  x \not \models \alpha$;\\
${\cal T},  x \models \alpha \wedge \beta$ & iff &
both ${\cal T},  x \models \alpha$ and
${\cal T},  x \models \beta$;\\
${\cal T},  x \models U(\alpha,\beta)$ & iff &
there is  $y>x$ in $T$ such that
${\cal T},  y \models \alpha$\\
&&and for all $z \in T$ such that
$x < z < y$ we have
${\cal T},  z \models \beta$; and\\
${\cal T},  x \models S(\alpha,\beta)$ & iff &
there is  $y<x$ in $T$ such that
${\cal T},  y \models \alpha$\\
&&and for all $z \in T$ such that
$y< z < x $ we have
${\cal T},  z \models \beta$.
\end{tabular}

Often, definitions, results or proofs
will have a {\em mirror image} 
in which $U$ and $S$ are exchanged
and $<$ and $>$ swapped.

A formula $\phi$ is {\em $\reals$-satisfiable}
if it has a real model:
i.e. there is a real structure
${\cal S}=(\reals,<,h)$
and $x \in \reals$
such that
${\cal S}, x\models \phi$.
A formula is {\em $\reals$-valid} iff 
it is true at all points
of all real structures.
Of course, a formula is $\reals$-valid
iff its negation is not
$\reals$-satisfiable.

Let RTL-SAT be the problem of
deciding whether a given formula
of $L(U,S)$ is $\reals$-satisfiable or
not. The main result
of this paper,
proved in
 lemma~\ref{lem:pspa}
and lemma~\ref{lem:phard} below,  is:

\begin{theorem}
RTL-SAT is PSPACE-complete.
\end{theorem}

\section{Mosaics for $U$ and $S$}
\label{sec:mosaic}

We will decide the satisfiability of formulas
by considering sets of small pieces of
real structures.
The idea is based on the mosaics
seen in \cite{Nem95}
and used in many other subsequent
proofs.

Each mosaic is a small piece of a model, i.e.
a small set of objects (points),
relations between them
and a set of formulas for each point indicating
which formulas are true
there in the whole model.
There will be {\em coherence} conditions
on the mosaic which are necessary for
it to be part of a larger model.

We want to show the equivalence of the existence of a model to
the existence of a certain set of mosaics:
enough mosaics to build a whole model.
So the whole set of mosaics also has to obey
some conditions. These are called {\em saturation}
conditions. For example, 
a particular small piece of a model might
require
a certain other piece to exist somewhere 
else in the model.
We talk of the first mosaic having a defect
which is cured by the latter mosaic.

Our mosaics will only be concerned with
a finite set of formulas:
\begin{definition}
For each
formula  $\phi$,
define the {\em closure} 
of $\phi$ to be
$\clos \phi= 
\{ \psi, \neg \psi \mid \psi \leq \phi \}$
where
$\chi \leq \psi$ means that $\chi$ is a subformula of
$\psi$.
\end{definition}
We can sometimes think of
$\clos \phi$ as being closed
under negation:
we could treat $\neg \neg \alpha$
as if it was $\alpha$.
To be more rigorous, we introduce the 
following notation.

\begin{definition}
For each $\alpha \in \lang$,
define
 $\sim \alpha$ to mean
$\beta$ if $\alpha = \neg \beta$
and
$\neg \alpha$ otherwise.
\end{definition}

Note that if $\alpha \in \clos \phi$ then
$\sim \alpha \in \clos \phi$.
Note also that in many places in the
proof we explicitly use $\neg \alpha$
when we can be sure it is in
$\clos \phi$,
for example when
$U(\alpha, \beta) \in \clos( \phi)$.

\begin{definition}
Suppose $\phi \in \lang$ and  $S \subseteq \clos \phi$.
Say $S$ is {\em propositionally consistent} (PC)
iff
there is no substitution instance of
a tautology of classical propositional logic
of the form $\neg (\alpha_1 \wedge ... \wedge \alpha_n)$ with
each $\alpha_i \in S$.
Say $S$ is {\em maximally propositionally consistent} (MPC)
iff
$S$ is maximal in being a subset of $\clos \phi$
which is PC. 
\end{definition}

We will define a mosaic to be
a triple $(A,B,C)$ of sets
of formulas.
The intuition is that
this corresponds to 
two points from a structure:
$A$ is the set of formulas
(from $\clos \phi$)
true at the earlier point,
$C$ is the set true at the
later point
and $B$ is the
set of formulas
which hold at all points
strictly in between.

\begin{definition}
Suppose $\phi$ is from $L(U,S)$.
A $\phi$-{\em mosaic} 
is a triple $(A,B,C)$ of 
subsets of $\clos \phi$
such that:\\
\begin{tabular}{ll}
0.1 & $A$ and $C$ are maximally propositionally consistent, and\\
0.2 & for all $\beta \in \clos(\phi)$ with $\neg \beta \in \clos(\phi)$
we have
 $\neg \beta \in B$ iff $\sim \beta \in B$\\
\end{tabular}\\ 
and the following four {\em coherency}
conditions hold:\\
\begin{tabular}{ll}
C1. & if $\neg U(\alpha,\beta)\in A$
and $\beta \in B$ then we have both:\\
& C1.1. $\neg \alpha \in C$ and 
either $\neg \beta \in C$
or $\neg U(\alpha,\beta) \in C$; and\\
& C1.2. $\neg \alpha \in B$ and $\neg U(\alpha,\beta) \in B$.\\
C2. & if $U(\alpha,\beta)\in A$
and $\neg \alpha \in B$ then
we have both:\\
& C2.1 either 
$\alpha \in C$
or both
$\beta \in C$ 
and $U(\alpha,\beta) \in C$; and\\
& C2.2. $\beta \in B$ and $U(\alpha,\beta) \in B$.\\
C3-4 & mirror images of C1-C2.
\end{tabular}
\end{definition}

\begin{definition}
If $m=(A,B,C)$ is a mosaic
then
$\start(m)=A$ is its {\em start},
$\cover(m)=B$ is its {\em cover}
and
$\fin(m)=C$ is its {\em end}.
\end{definition}

If we start to build a model using mosaics
then, as we have noted, we may realise that
the inclusion of one mosaic
necessitates the inclusion of others:
defects need curing.

\begin{definition}
A {\em defect} in a mosaic $(A,B,C)$
is either
\begin{tabbing}
1. \= a formula $U(\alpha,\beta) \in A$
with either\\
\> 1.1 \= $\beta \not \in B$,\\
\> 1.2 \> ($\alpha \not \in C$
and $\beta \not \in C$), or\\
\> 1.3 \> ($\alpha \not \in C$
and $U(\alpha,\beta) \not \in C$); \\
2. \>  a formula $S(\alpha,\beta) \in C$
with either\\
\> 2.1 \> $\beta \not \in B$,\\
\> 2.2 \> ($\alpha \not \in A$
and $\beta \not \in A$), or\\
\> 2.3 \> ($\alpha \not \in A$
and $S(\alpha,\beta) \not \in A$); or\\
3. \> a formula $\beta \in \clos \phi$
with $\sim \beta \not \in B$.
\end{tabbing}
\end{definition}

We refer to defects of type 1 to 3 (as listed here).
Note that the same formula may be both 
a type 1 or 2 defect and a type 3 defect
in the same mosaic. In that case
we count it as two separate defects.

We will need to string mosaics
together to build
linear orders.
This can only be done under certain conditions.
Here we introduce the idea
of composition
of mosaics.

\begin{definition}
We say that $\phi$-mosaics
$(A',B',C')$ and $(A'',B'',C'')$
{\em compose}
iff
$C'=A''$.
In that case,
their {\em composition}
is $(A',B' \cap C' \cap B'',C'')$.
\end{definition}
It is straightforward to
prove that this is a mosaic
and that composition
of mosaics is associative.

\begin{nemma}
{compos}
If mosaics $m$ and $m'$ compose
then their composition
is a mosaic.
\end{nemma}

\sfproof

\begin{nemma}
{assoc}
Composition of mosaics is associative.
\end{nemma}

\sfproof

Thus we can talk of
sequences
of mosaics composing
and then find their
composition.
We define the composition of 
a sequence of length one to be
just the mosaic itself.
We leave the composition
of an empty sequence undefined.

\begin{definition}
A {\em decomposition}
for a mosaic $(A,B,C)$
is any finite
sequence of mosaics
\(
(A_1,B_1,C_1), (A_2,B_2,C_2), ...,
(A_n,B_n,C_n)
\)
which
composes to $(A,B,C)$.
\end{definition}

It will be useful to introduce
an idea of fullness of decompositions.
This is intended to 
be a decomposition which provides witnesses
to the cure of every defect in the
decomposed mosaic.

\begin{definition}
The decomposition above
is {\em full}
iff the following
three conditions all hold:
\begin{tabbing}
1. \= for all $U(\alpha,\beta) \in A$
we have\\
\> 1.1. \= $\beta \in B$ and either
($\beta \in C$ and $U(\alpha,\beta) \in C$)
or
$\alpha \in C$,\\
\> 1.2. \> or there is some $i$
such that $1\leq i < n$,
$\alpha \in C_i$,
for all $j \leq i$,
$\beta \in B_j$\\
\> \> and 
for all $j<i$, $\beta \in C_j$;\\
2. \> the mirror image of 1.; and\\
3. \> for each $\beta \in \clos \phi$
such that $\sim \beta \not \in B$
there is some $i$ such that
$1 \leq i < n$\\
\> and
$\beta \in C_i$.
\end{tabbing}
\end{definition}

If 1.2 above holds 
in the case that $U(\alpha,\beta) \in A$
is a type 1 defect in $(A,B,C)$
then we say that
{\em a cure for the defect is witnessed}
(in the decomposition) 
by the end of $(A_i,B_i,C_i)$
(or equivalently by the start of $(A_{i+1},B_{i+1},C_{i+1})$).
Similarly for the mirror image for $S(\alpha,\beta) \in C$.
If $\beta \in C_i$ is a type 3 defect
in $(A,B,C)$ then we also say that
{\em a cure for this defect is witnessed}
(in the decomposition) 
by the end of $(A_i,B_i,C_i)$.
If a cure for any defect is witnessed
then we say that the defect
is cured.

\begin{nemma}
{fullcure}
If $m_1, ..., m_n$ is a full decomposition of
$m$ then every defect in $m$
is cured in the decomposition.
\end{nemma}

\sfproof

\section{Satisfiability and relativization}

Because mosaics represent linear orders
with end points, it is
inconvenient for us to 
continue to work directly with $\reals$.
Because we want to make use
of some simple tricks with
the metric at several places
in the proof, we will
move to work in the unit interval
$\unit$ instead.

If $x<y$ from $\reals$ then let 
$]x,y[$ denote the open interval
$\{ z\in \reals | x<z<y \}$
and 
$[x,y]$ denote the closed interval
$\{ z \in \reals | x \leq z \leq y \}$.
Similarly for half open intervals.

One can get a mosaic from any two points
in a structure.

\begin{definition}
If  ${\cal T}=(T,<,h)$ is a 
structure
and $\phi$  a formula 
then for each $x<y$ from $T$ we define
$\mos^\phi_{\cal T}(x,y)= (A,B,C)$ where:\\
\begin{tabular}{ll}
$A = $ & $ \{ \alpha \in \clos \phi |
{\cal T}, x \models \alpha \}$,\\
$B = $ & $\{ \beta \in \clos \phi |
\mbox{ for all }
z \in T,
\mbox{ if }
x<z<y
\mbox{ then } {\cal T}, z \models \beta \}$, and\\
 $C = $ & $\{ \gamma \in \clos \phi |
{\cal T}, y \models \gamma \}$.\\
\end{tabular}
\end{definition}

It is straightforward to show that this is a mosaic.

\begin{nemma}
{mosaic}
$\mos^{\phi}_{\cal T}(x,y)$
is a mosaic.
\end{nemma}

\sfproof

If $\cal T$ and $\phi$ are
clear from context
then we just write
$\mos(x,y)$ for
$\mos^\phi_{\cal T}(x,y)$.

\begin{definition}
Suppose $T \subseteq \reals$.
Let $<$ also 
denote the restriction of $<$
to any such $T$.
We say that a $\phi$-mosaic is {\em
$T$-satisfiable} iff it is $\mos^\phi_{\cal
T}(x,y)$ for some $x<y$ from $T$
and some  structure ${\cal
T}=(T,<,h)$. 
\end{definition}

\begin{definition}
We say that a $\phi$-mosaic is {\em fully $\unit$-satisfiable}
iff it is $\mos^\phi_{\cal T}(0,1)$
from some  structure ${\cal T}=(\unit,<,h)$.
\end{definition}

We will now relate the
satisfiability of a formula
$\phi$ to
that of certain mosaics.

\begin{definition}
Given $\phi$
and an atom $q$ which does
not appear in $\phi$,
we define a map $* = *_q^{\phi}$ 
on formulas in $\clos(\phi)$ recursively:\\
\begin{tabular}{ll}
1. & $* p = p \wedge q$,\\
%2. & $* \truth = q$,\\
2. & $* \neg \alpha = \neg ( * \alpha ) \wedge q$,\\
3. & $* ( \alpha \wedge \beta )=
* (\alpha ) \wedge * ( \beta ) \wedge q$,\\
4. & $* U (\alpha, \beta) =
U( * \alpha, * \beta ) \wedge q$, and\\
5. & $* S (\alpha, \beta) =
S( * \alpha, * \beta ) \wedge q$.\\
\end{tabular}
\end{definition}

So $*_q^{\phi}(\phi)$
will be a formula using only $q$ and atoms from
$\phi$.

\begin{nemma}
{threetimes}
$*_q^{\phi}(\phi)$
 is at most 3 times as long as $\phi$. 
\end{nemma}

\sfproof

\begin{nemma}
{inclosstar}
If $\alpha \leq \phi$ then
$* \alpha \leq * \phi$.
\end{nemma}

\sfproof

\begin{definition}
We say that a $*_q^{\phi}(\phi)$-mosaic $(A,B,C)$
is $(\phi,q)$-{\em relativized}
iff\\
\begin{tabular}{ll}
1. & $\neg q$ is in $A$ and no $S(\alpha,\beta)$ is in $A$;\\
2. & $q \in B$ and $\neg *_q^{\phi}( \phi) \not \in B$; and\\
3. & $\neg q \in C$ and no $U(\alpha,\beta)$ is in $C$.\\
\end{tabular}
\end{definition}

\begin{nemma}{satfmos}
Suppose that $\phi$ is a formula of $L(U,S)$
and $q$ is an atom not appearing
in $\phi$.
Then $\phi$ is $\reals$-satisfiable
iff there is 
 some fully $\unit$-satisfiable
$(\phi,q)$-relativized 
$*_q^{\phi}(\phi)$-mosaic.
\end{nemma}

\begin{proof}
Let $*=*_q^{\phi}$ and let $\zeta: ]0,1[ \rightarrow \reals
$ be any order preserving bijection.

Suppose that $\phi$ is $\reals$-satisfiable.
Say that
${\cal S}=(\reals,<,g)$, $s_0 \in \reals$
and ${\cal S}, s_0 \models \phi$.
Let
${\cal T}=(\unit,<,h)$ where:\\
\begin{tabular}{ll}
1. & for atom $p \neq q$, $h(p)= 
\{ t \in ]0,1[ | \zeta(t) \in g(p) \}$;
and\\
2. & $h(q)= ]0,1[$.\\
\end{tabular}\\
An easy induction on the construction
of formulas in $\clos(*\phi))$ shows that
${\cal T}, \zeta^{-1}(s_0) \models * \phi$
and
so $\mos^{*\phi}_{\cal T}(0,1)$
is the right mosaic.

Suppose mosaic $(A,B,C)=\mos(0,1)$
from structure ${\cal T}=(\unit,<,h)$
is a $(\phi,q)$-relativized 
$*(\phi)$-mosaic.
Thus
$q \in B$ and $\neg q \in A \cap C$.
Define 
${\cal S}= (\reals,<,g)$ via
 $s \in g(p)$ iff $\zeta^{-1}(s) \in h(p)$
for any atom $p$ (including $p=q$).
As $\neg * \phi \not \in B$
there is some $z$ such that
$0<z<1$ and
${\cal T},z \models * \phi$.
It is easy to show that
${\cal S}, \zeta(z) \models \phi$.
\end{proof}

Our satisfiability procedure will
be to guess
a relativized mosaic $(A,B,C)$
and then check  that
$(A,B,C)$
is fully $\unit$-satisfiable.
Thus we now turn
to the question
of deciding
whether a relativized mosaic
is satisfiable.

\section{Shuffles}

A game can be played by two players
with mosaics:
one player provides full decompositions
for the mosaics chosen by the other.
We will not develop this idea here
but we will examine some
structures which correspond
to tactics in this game.
In this section we will consider
the most complex such structure:
the shuffle.

We shall write
$\langle p_1, ..., p_n \rangle$
for the sequence of mosaics
containing $p_1, ..., p_n$
in that order.
We shall write
$\pi \cat \rho$
for the sequence resulting
from the concatenation
of sequences $\pi$ and $\rho$
in that order.
Sequences will always be finite.

\begin{definition}
Suppose $0 \leq r$,
each $\lambda_i (1 \leq i \leq r)$ is a
non-empty composing sequence of $\phi$-mosaics,
and $P_0, ..., P_s$ $(0 \leq s)$ are maximally propositionally consistent
subsets of $\clos \phi$.

Suppose $\phi$-mosaic $o= (A,B,C)$
and:\\
\begin{tabular}{l}
$m' = ( A, B , P_0 )$;\\
$y_i= (P_i,B,P_{i+1}) \; (0 \leq i \leq s-1)$;\\
$y_s=( P_s,B,P_0)$;\\
$m'' = ( P_0, B ,C)$; and\\
$\mu = \langle y_0, ..., y_s \rangle$.\\
\end{tabular}\\

If $r=0$
suppose $\lambda= \langle  \rangle$, the empty sequence,
but otherwise, if $r>0$, suppose:\\
\begin{tabular}{l}
$A_i$ is the start of the first mosaic in
$\lambda_{i}(1 \leq i \leq r)$;\\ 
$C_i$ is the end of the last mosaic
in $\lambda_{i}(1 \leq i \leq r)$;\\ 
$x_0 = ( P_0, B , A_1)$;\\
$x_i = ( C_{i}, B  , A_{i+1} )$, $(1 \leq i \leq r-1)$;\\
$x_{r} = ( C_r, B, P_0)$;\\
$\lambda= \langle x_0 \rangle \cat \lambda_1 \cat
\langle x_1 \rangle \cat ...
\cat \lambda_r \cat \langle x_r \rangle$.\\
\end{tabular}\\

Further suppose that $m'$, $m''$, and each $y_i$ and $x_i$
are mosaics.

Then we say that $o$ 
is {\em fully decomposed by
the tactic
shuffle
$( \langle P_0, ..., P_s \rangle,
\langle \lambda_{1},
...,
\lambda_{r}
\rangle )$}
iff the following conditions all hold:\\
\begin{tabular}{ll}
F1. & $o$ is fully decomposed by
$\langle m' \rangle \cat \lambda \cat \mu \cat 
\langle m'' \rangle$;\\
F2. & if $r>0$, $x_0$ is fully decomposed by
$\lambda \cat \mu \cat 
\langle x_0 \rangle$;\\
F3. & if $0<i<r$, $x_i$ is fully decomposed by\\
& $\langle x_i \rangle \cat 
\lambda_{i+1} \cat \langle x_{i+1} \rangle \cat
... \cat \lambda_r \cat
\langle x_r \rangle \cat
\mu \cat \langle x_0 \rangle \cat
\lambda_1 \cat \langle x_1 \rangle \cat ...
\cat \lambda_i \cat \langle x_i \rangle$;\\
F4. & if $r>0$, $x_r$ is fully decomposed by
$\langle x_r \rangle \cat \mu \cat \lambda$;\\
F5. & if $0 \leq i < s$, $y_i$ is fully decomposed by
$\langle y_i, y_{i+1}, ..., y_s \rangle 
\cat \lambda \cat
  \langle y_0, ..., y_i \rangle$;\\
F6. & $y_s$ is fully decomposed by
$\langle y_s \rangle \cat \lambda \cat \mu$.\\
\end{tabular}\\
\end{definition}

The term shuffle has been used  
in the literature (see, for example, \cite{LaL66} or
\cite{BuG85})
to refer to a certain method of 
constructing a monadic linear structure
from a thorough mixture
of smaller linear structures.
The intention here is similar.

Note that as $s \geq 0$ there is at least
one $P_i$ involved in the shuffle.
In a general linear order setting we could 
define a shuffle with no $P_i$s (provided
that then $r>0$) but over the reals it turns out
to be crucial to require at least
one $P_i$.
This ensures that the mosaic is
satisfiable in a structure
on a separable linear frame.

For the purposes of
algorithmic checking of shuffles
we find it convenient to have a 
different characterization
of shuffles.
First a couple of helpful properties.

\begin{definition}
Suppose $\phi \in \lang$ and
$m$ is a $\phi$-mosaic.
We say that an MPC set $Q \subseteq \clos(\phi)$
satisfies the forward $K(m)$ property
iff
for any $U(\alpha, \beta) \in \clos(\phi)$
we have
$U(\alpha, \beta) \in Q$ iff
both $\beta \in \cover(m)$ and
(at least) one of the following holds:\\
\begin{tabular}{ll}
K1 & $\sim \alpha \not \in \cover(m)$;\\
K2 & $\alpha \in \fin(m)$; or\\
K3 & $\beta \in \fin(m)$ and $U(\alpha,\beta) \in \fin(m)$.\\
\end{tabular}\\
The mirror image is the backwards $K(m)$ property.
\end{definition}

\begin{nemma}
{shufcon}
Suppose $\phi \in \lang$,
$m=(A,B,C)$ is a $\phi$-mosaic,
and each $P_i \subseteq \clos(\phi)\; (0 \leq i \leq s)$
and each $\lambda_i \;(1 \leq i \leq r)$ is
a sequence of $\phi$-mosaics.

Then $m$ is fully decomposed by the tactic
shuffle $(\langle P_0, ..., P_s \rangle,
\langle \lambda_1, ..., \lambda_r \rangle )$
iff the following seven conditions hold:\\
\begin{tabular}{ll}
S0 &
$B$ is a subset
of each $P_i$
and of the start, end and cover of each mosaic
in each $\lambda_i$;\\
S1 &
each $P_i$ 
satisfies both the forward and backwards 
$K(m)$ property;\\
S2 &
the start of the first mosaic in each $\lambda_i$ 
satisfies the backwards $K(m)$ property;\\
S3 &
the end of the last mosaic in each $\lambda_i$ 
satisfies the forwards $K(m)$ property; \\
S4 &
$A$ satisfies the forward $K(m)$ property;\\
S5 &
$C$ satisfies the backwards $K(m)$ property;\\
S6 &
if $\beta \in \clos(\phi)$
but $\sim \beta \not \in B$ then
either $\beta$ is contained in some $P_i$
or\\
& $\beta$ is contained in the start or end
of some mosaic in some $\lambda_i$.
\end{tabular}
\end{nemma}

\begin{proof}

Consider the forward direction of the proof.
Suppose $m=(A,B,C)$ is fully decomposed by the tactic
shuffle $(\langle P_0, ..., P_s \rangle,
\langle \lambda_1, ..., \lambda_r \rangle )$.

By F0, we have a decomposition for $m$ including each $\lambda_i$
and mosaics with each $P_i$ in their starts
or ends.
$S0$ follows.

We now establish condition S1.
Each $P_i$ is an MPC by the definition of a shuffle.
To show the forward $K(m)$ property for $P_i$,
suppose that $U(\alpha,\beta) \in P_i$.
We consider the case when $i<s$:
the case with $i=s$ is similar.
We know (F4-F5) that 
$y_i = (P_i,B,P_{i+1})$ is fully decomposed by
$\langle y_i, ..., y_s \rangle \cat \lambda
\cat \langle y_0, ..., y_i \rangle$.
If $U(\alpha,\beta)$ is a type 1 defect
in $y_i$ then it is cured in this
decomposition and we can conclude that
$\beta$ is in the cover $B$ of the first mosaic $y_i$.
If $U(\alpha,\beta)$ is not a type 1 defect
in $y_i$ then $\beta \in \cover(y_i)=B$ as well.
Thus in any case $\beta \in B$.

I claim that $U(\alpha, \beta) \in A$.
If not then
$\neg U(\alpha,\beta) \in \start(m)$ and coherency C1.2 of $m$
implies that
$\neg U (\alpha, \beta) \in B \subseteq P_i$.
This is a contradiction to the consistency of $P_i$.

So $U(\alpha, \beta) \in A$ is either a type 1 defect
in $m$ or not.
In the former case it is cured in the full decomposition ($F1$)
for $m$ and so $\alpha$ appears in the start or end of
a mosaic in some $\lambda_j$ or in some $P_j$.
Thus $\neg \alpha \not \in \cover(m)=B$.
This is K1.

If  $U(\alpha, \beta) \in A$ is not a type 1 defect
in $m$
then K2 or K3 holds by definition.

We now show the converse part of the
forward $K(m)$ property for $P_i$.
Suppose that both
$\beta \in \cover(m)$
and K1 holds:
the cases of K2 or K3 holding are straightforward.
Thus $\sim \alpha \not \in B$,
 $\alpha$ is a type 3 defect
in $m$
and so a cure
is witnessed in some $P_j$
or in the start or end of some
mosaic in some $\lambda_j$.
Now look in the decomposition
F5 (or F6) for $y_i$
in which we have $\beta$ holding
in all starts, ends and covers and
$\alpha$ appearing somewhere.
A simple induction shows
that we must have
$U(\alpha, \beta)$ in the
very start $P_i$
as required.

To show the backwards $K(m)$ property is the mirror image.

Very similar arguments establish conditions
S2 -- S6.
To show condition S2
we just use
the full decomposition (F2-F3) 
for $x_{i-1}$ and reason about type 2 defects.
To show condition S3,
use the full decomposition (F2-F4) for
$x_i$ and reason about type 1 defects.
Conditions S4, S5 and S6 follow from using
the full decomposition (F1)  for $m$
and reasoning about type 1, 2 and 3 defects respectively.

Now consider the converse: suppose
that the seven conditions S0--S6 hold
for mosaic $m=(A,B,C)$.

First we must show that each of
$m'$, $m''$, each $y_i$ and any $x_i$ 
(as defined from $m$, the $P_i$ and the $\lambda_i$
 in the definition of a shuffle) are mosaics.
This follows from
\begin{claim}
If the MPC $D \subseteq \clos(\phi)$
satisfies the forward $K(m)$ property,
the MPC $E \subseteq \clos(\phi)$ satisfies the
backwards $K(m)$ property
and $B \subseteq D \cap E$
then
$(D,B,E)$ is a mosaic.
\end{claim}

\begin{proof}
We must check the first two coherency conditions.
The mirror images are mirror images.

(C1). Suppose $\neg U(\alpha,\beta)\in D$
and $\beta \in B$.

First we 
establish that
we must have $U(\alpha, \beta) \not \in A$.
Suppose not for contradiction.
Since $U(\alpha,\beta) \not \in D$,
K1 does not hold and so $\neg \alpha \in B$,
K2 does not hold and so $\neg \alpha \in C$
and
K3 does not hold and so either $\beta \not \in C$
or $U(\alpha,\beta) \not \in C$. 
We have a contradiction
to the coherency (C2.1) of $m$.

(C1.1). First, we show $\neg \alpha \in E$.
Otherwise, $\neg \alpha \not \in B \subseteq E$.
Thus, by K1, $U(\alpha,\beta) \in D$
and we have our contradiction.

Next we show that either $\neg \beta \in E$
or $\neg U(\alpha,\beta) \in E$.
Suppose instead that $\beta \in E$ and
$U(\alpha,\beta) \in E \supseteq B$.
Thus $\neg U(\alpha,\beta) \not \in B$.
By coherency C1.2 of $(A,B,C)$,
we must have $U(\alpha,\beta) \in A$
which is a contradiction.

(C1.2). We show that $\neg \alpha \in B$ and $\neg U(\alpha,\beta) \in B$.
We can not have $\neg \alpha \not \in B$,
as then K1 implies
$U(\alpha, \beta) \in D$.
We can not have $\neg U(\alpha,\beta) \not \in B$
as then coherency (C1.2) of $m$ implies $U(\alpha,\beta) \in A$,
a contradiction.

(C2).
Assume $U(\alpha,\beta) \in D$ and $\neg \alpha \in B$.
By the forward $K(m)$ property for $D$,
$\beta \in B \subseteq E$ and, since $\neg \alpha \in B$,
either K2    or K3 holds (for $C$).
By the coherency C1.1 of $m$ we can conclude
that we can not have $\neg U(\alpha, \beta) \in A$.
Thus $U(\alpha, \beta) \in A$
and C2.2 of $m$ implies
that
$U(\alpha, \beta) \in B \subseteq E$
as required.
\end{proof}

Next we must check the fullness of the
decompositions.
This follows by
\begin{claim}
Suppose $D$ and $E$ are as in the previous claim.

Furthermore, suppose the sequence $\sigma$ of mosaics
composes to $(D,B,E)$
such that
for each $\beta \in \clos(\phi)$ with $\sim \beta \not \in B$,
there is a mosaic in $\sigma$
other than the very first
which includes $\beta$ in its start.

Then $(D,B,E)$ is fully decomposed
by $\sigma$.
\end{claim}

\begin{proof}
{\bf Type 1 defects:}
Suppose $U(\alpha,\beta)\in D$ 
is a type 1 defect
of $(D,B,E)$.
By the forward $K(m)$ property for $D$,
$\beta \in B$.

As $U(\alpha, \beta)$ is a type 1 defect 
$\alpha \not \in E$
and either
$\beta \not \in E$
or
$U(\alpha,\beta) \not \in E$.
By coherency C2 of
$(D,B,E)$, $\neg \alpha \not \in B$.
So $\sim \alpha \not \in B$
and $\alpha$ must appear in a non-first
mosaic in $\sigma$ and we have our cure.

{\bf Type 2 defects:} mirror image.

{\bf Type 3 defects:}
Suppose $\beta \in \clos( \phi)$
but $\sim \beta \not \in B$.
Thus $\beta$ appears in the
start of a non-first mosaic in $\sigma$.
We have our witness.
\end{proof}

Thus 
$m$ is fully decomposed by the tactic
shuffle $(\langle P_0, ..., P_s \rangle,
\langle \lambda_1, ..., \lambda_r \rangle )$
as required.
\end{proof}

\section{Real Mosaic Systems}
\label{sec:rms}

In this section we define
a concept of a collection or system
of mosaics in which
each member is decomposable
in terms of simpler members.
First another tactic for
decomposition.

\begin{definition}
Suppose $\phi \in \lang$,
$m$ is a $\phi$-mosaic and $\sigma$
is a non-empty sequence
of $\phi$-mosaics.
Then, we say that $m$ is fully decomposed by
the {\em tactic} 
$\lead( \sigma )$ iff
$\langle m \rangle \cat \sigma$
is a full decomposition of $m$.
We say that $m$ is fully decomposed by
the {\em tactic}
$\trail( \sigma )$ iff
$\sigma \cat \langle m \rangle$
is a full decomposition of $m$.
\end{definition}

\begin{definition}
For $\phi \in \lang$,
suppose $S$
is a set of $\phi$-mosaics
and $n \geq 0$.

A $\phi$-mosaic $m$ 
is a {\em level $n^{+}$ 
member of $S$}
iff $m$ is the composition
of a sequence of mosaics, each of them
being either
a level $n$ member of $S$ or
fully decomposed by the tactics
$\lead( \sigma)$ or $\trail(\sigma)$ with each mosaic in $\sigma$
being a level $n$ member of $S$.

A $\phi$-mosaic $m$ 
is a {\em level $(n+1)^{-}$ 
member of $S$}
iff $m$ is the composition
of a sequence of mosaics, each of them
being either
a level $n^+$ member of $S$ or
fully decomposed by the tactics
$\lead( \sigma)$ or $\trail(\sigma)$ with each mosaic in $\sigma$
being a level $n^+$ member of $S$.

A $\phi$-mosaic $m \in S$ is a
{\em level $n$ member of $S$}
iff $m$ is the composition
of a sequence of mosaics with each 
of them being either
a level $n^-$ 
member of $S$ or
a mosaic which is fully decomposed by the tactic
$\shuffle( \langle P_0 , ..., P_s \rangle , 
\langle \sigma_1, ..., \sigma_r \rangle)$
with each mosaic in each $\sigma_i$
being a level $n^-$ 
member of $S$.
\end{definition}

Note that it is 
generally possible
for mosaics to be level 0 
members of some $S$
provided that they are compositions of
mosaics which can be fully decomposed by shuffles in which there
are no sequences (ie, $r=0$).

\begin{definition}
For $\phi \in \lang$,
a {\em real mosaic system} of $\phi$-mosaics
is a set $S$ of $\phi$-mosaics 
such that for every $m \in S$
there exists some $n$ such that $m$
is a level $n$ member of $S$.
For any $n$ we say that $S$ is a real mosaic system
of depth $n$ iff every $m \in S$
is a level $n$ member of $S$.
\end{definition}

\section{Realizing Mosaics}
\label{sec:realize}

In this section we show that relativized
mosaics which appear in real mosaic
systems are satisfiable.
To do so we define a concept of realization
intended to capture the idea
of a mosaic being satisfiable as
far as internal information is
concerned:
ie we ignore formulas of the
form $U(\alpha,\beta)$ in the end
or $S(\alpha,\beta)$ in the start.

\begin{definition}
Suppose that $x<y$ from $\unit$.
We say that $\phi$-mosaic $m$
is realised by the
map $\mu$
on the closed interval $[x,y]$
iff the following conditions all hold:
\begin{tabbing}
R1. \= for each $z \in [x,y]$, $\mu(z)$ is
a maximally propositionally consistent
subset of $\clos \phi$;\\
R2. \> Suppose $z \in [x,y[$. Then $U(\alpha,\beta) \in \mu(z)$
iff either\\
\> R2.1,  \= there is $u$ such that
$z< u \leq y$ and $\alpha \in \mu(u)$
and for all $v$,\\
\> \> if $z < v <u$ then $\beta \in \mu(v)$
or\\
\> R2.2, \> $\beta \in \mu(y)$, $U(\alpha, \beta) \in \mu(y)$
and
 for all $v$,
if $z < v <y$ then $\beta \in \mu(v)$;\\
R3. \> the mirror image of R2 for $S(\alpha,\beta)$;\\
R4. \> $\mu(x)$ is the start of $m$;\\
R5. \> $\mu(y)$ is the end of $m$; and\\
R6. \> for each $\beta \in \clos \phi$,
$\beta$ is in the cover of $m$
iff
for all $u$,
if $x<u<y$, $\beta \in \mu(u)$.
\end{tabbing}
\end{definition}

\begin{nemma}
{realcompo}
If $m$ is the composition of $m'$ and $m''$
with each of $m'$ and $m''$
having a realization on any closed interval of $\unit$
then for any $x<y$ from $\unit$,
there is $\mu$ which realises
$m$ on $[x,y]$.
\end{nemma}

\begin{proof}
Given $x<y$ from $\unit$ choose any $w$ with $x<w<y$.
Let $\mu'$ realize $m'$ on $[x,w]$ and
$\mu''$ realize $m''$ on $[w,y]$.
Define $\mu: [x,y] \rightarrow \power {\clos \phi}$ via:
\[
\mu(u) = \left \{
\begin{array}{ll}
\mu'(u), & x \leq u \leq w\\
\mu''(u), & w < u \leq y\\
\end{array}
\right .
\]

It is straightforward to check that
$\mu$ realizes $m$ on $[x,y]$.
Use the facts that
$\fin(m')=\start(m'')$ and
$\cover(m) = \cover(m') \cap \start(m'') \cap \cover(m'')$.

\end{proof}

\begin{nemma}
{lead}
If $m$ is fully decomposed by the tactic
$\lead( \sigma)$ with each mosaic in $\sigma$
having a realization on any closed interval of $\unit$
then for any $x<y$ from $\unit$,
there is $\mu$ which realises
$m$ on $[x,y]$.
There is a mirror image result for $\trail(\sigma)$.
\end{nemma}

\begin{proof}
Say $\sigma= \langle b_1, ..., b_k \rangle$.
Choose a sequence $x < ... < y_2 < y_1 < y_0=y$
converging to $x$.
For each $i=1,2, ...$, and each $j \in J=\{1,...,k\}$,
let
$\mu_{i,j}$ realize $b_j$
on $[y_{ik-j+1}, y_{ik-j}]$.

\begin{figure}
\begin{center}

\setlength{\unitlength}{3947sp}%
\begingroup\makeatletter\ifx\SetFigFont\undefined%
\gdef\SetFigFont#1#2#3#4#5{%
  \reset@font\fontsize{#1}{#2pt}%
  \fontfamily{#3}\fontseries{#4}\fontshape{#5}%
  \selectfont}%
\fi\endgroup%
\begin{picture}(5649,966)(1264,-4540)
\thinlines
\put(1276,-4036){\line( 1, 0){5625}}
\put(6676,-3661){\line( 0,-1){600}}
\put(1426,-3661){\line( 0,-1){600}}
\put(5851,-3886){\line( 0,-1){300}}
\put(5251,-3886){\line( 0,-1){300}}
\put(4651,-3886){\line( 0,-1){300}}
\put(4126,-3886){\line( 0,-1){300}}
\put(3676,-3886){\line( 0,-1){300}}
\put(3226,-3886){\line( 0,-1){300}}
\put(2831,-3886){\line( 0,-1){300}}
\put(2476,-3886){\line( 0,-1){300}}
\put(2176,-3886){\line( 0,-1){300}}
\put(1951,-3886){\line( 0,-1){300}}
\put(1876,-3886){\line( 0,-1){300}}
\put(1801,-3886){\line( 0,-1){300}}
\put(1726,-3886){\line( 0,-1){300}}
\put(6151,-3961){\makebox(0,0)[lb]{\smash{\SetFigFont{12}{14.4}{\rmdefault}{\mddefault}{\updefault}$b_k$}}}
\put(5401,-3961){\makebox(0,0)[lb]{\smash{\SetFigFont{12}{14.4}{\rmdefault}{\mddefault}{\updefault}$b_{k-1}$}}}
\put(4801,-3961){\makebox(0,0)[lb]{\smash{\SetFigFont{12}{14.4}{\rmdefault}{\mddefault}{\updefault}$...$}}}
\put(4276,-3961){\makebox(0,0)[lb]{\smash{\SetFigFont{12}{14.4}{\rmdefault}{\mddefault}{\updefault}$b_2$}}}
\put(3826,-3961){\makebox(0,0)[lb]{\smash{\SetFigFont{12}{14.4}{\rmdefault}{\mddefault}{\updefault}$b_1$}}}
\put(3376,-3961){\makebox(0,0)[lb]{\smash{\SetFigFont{12}{14.4}{\rmdefault}{\mddefault}{\updefault}$b_k$}}}
\put(2866,-3961){\makebox(0,0)[lb]{\smash{\SetFigFont{12}{14.4}{\rmdefault}{\mddefault}{\updefault}$b_{k-1}$}}}
\put(2626,-3961){\makebox(0,0)[lb]{\smash{\SetFigFont{12}{14.4}{\rmdefault}{\mddefault}{\updefault}$...$}}}
\put(1501,-4186){\makebox(0,0)[lb]{\smash{\SetFigFont{12}{14.4}{\rmdefault}{\mddefault}{\updefault}$...$}}}
\put(1276,-4411){\makebox(0,0)[lb]{\smash{\SetFigFont{12}{14.4}{\rmdefault}{\mddefault}{\updefault}$x$}}}
\put(5776,-4411){\makebox(0,0)[lb]{\smash{\SetFigFont{12}{14.4}{\rmdefault}{\mddefault}{\updefault}$y_1$}}}
\put(5201,-4411){\makebox(0,0)[lb]{\smash{\SetFigFont{12}{14.4}{\rmdefault}{\mddefault}{\updefault}$y_2$}}}
\put(4901,-4411){\makebox(0,0)[lb]{\smash{\SetFigFont{12}{14.4}{\rmdefault}{\mddefault}{\updefault}$...$}}}
\put(4476,-4411){\makebox(0,0)[lb]{\smash{\SetFigFont{12}{14.4}{\rmdefault}{\mddefault}{\updefault}$y_{k-2}$}}}
\put(4051,-4411){\makebox(0,0)[lb]{\smash{\SetFigFont{12}{14.4}{\rmdefault}{\mddefault}{\updefault}$y_{k-1}$}}}
\put(3601,-4411){\makebox(0,0)[lb]{\smash{\SetFigFont{12}{14.4}{\rmdefault}{\mddefault}{\updefault}$y_k$}}}
\put(3151,-4411){\makebox(0,0)[lb]{\smash{\SetFigFont{12}{14.4}{\rmdefault}{\mddefault}{\updefault}$y_{k+1}$}}}
\put(2701,-4411){\makebox(0,0)[lb]{\smash{\SetFigFont{12}{14.4}{\rmdefault}{\mddefault}{\updefault}$y_{k+2}$}}}
\put(2176,-4411){\makebox(0,0)[lb]{\smash{\SetFigFont{12}{14.4}{\rmdefault}{\mddefault}{\updefault}$...$}}}
\put(6601,-4411){\makebox(0,0)[lb]{\smash{\SetFigFont{12}{14.4}{\rmdefault}{\mddefault}{\updefault}$y_0=y$}}}
\end{picture}
\end{center}
\caption{Realizing lead tactics}
\end{figure}
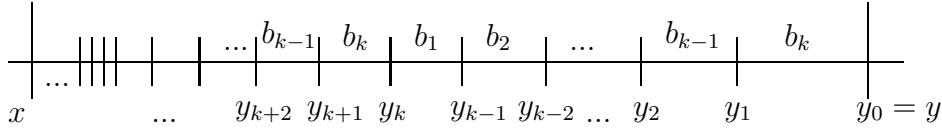

Define $\mu: [x,y] \rightarrow \power {\clos \phi}$
via 
$\mu(x)= \start(m), \mu(y)=\fin(m)=\fin(b_k)=\start(b_1)$
and if $z \in ]y_{ik-j+1}, y_{ik-j}]$,
then put $\mu(z)= \mu_{i,j}(z)$.

I claim that $\mu$ realizes $m$ on $[x,y]$.
Consider the six realization conditions.
The harder cases are conditions R2 and R6.
There are several subcases and their converses
and they all involve similar sorts of reasoning
so we will just present a few for illustration
purposes.

To show the forward direction of R2
assume that $z \in [x,y[$ and $U(\alpha,\beta) \in \mu(z)$.
The subcases concern whether
$z=x$, $z$ equals some $y_{ik-j}<y_0$
or $z$ is in some $]y_{ik-j+1},y_{ik-j}[$.
We must show that R2.1 or R2.2 holds.

Suppose $x=z$ and $U(\alpha,\beta) \in \mu(x)$ is a type 1 defect
in $m$. 
As $\langle m, b_1, ..., b_k \rangle$
is a full decomposition of $m$,
a cure to this defect is witnessed in this
sequence.
We can conclude that
$\beta$ is in the cover of $m$ and 
so in the starts, covers and ends of each of the $b_i$.
We can also conclude that
$\alpha$ is in the start of some $b_i$
and in the end of the preceding one $b_j$
(with $j=k$ if $i=1$).
R2.1 follows easily with $u=y_{k-j}$.

Suppose $x=z$ and $U(\alpha,\beta)  \in \mu(x)$ is not a type 1 defect
in $m$.
So $\beta \in \cover(m)$.
If $\alpha \in \fin(m)$ then R2.1 holds.
Otherwise R2.2 holds.

Suppose  $U(\alpha,\beta) \in \mu(z)$
and $z \in ]y_{ik-j+1},y_{ik-j}[$.
So $U(\alpha,\beta) \in \mu_{i,j}(z)$.
Now $b_j$ is realized by
$\mu_{i,j}$ on $[y_{ik-j+1},y_{ik-j}]$
and so by R2 (for $\mu_{i,j}$)
either R2.1 holds and we are almost immediately done
or R2.2 holds.
In this latter case
$\beta \in \fin(b_j)$,
$U(\alpha,\beta) \in \fin(b_j)$,
we may suppose $\neg \alpha \in \fin(b_j)$
and
for all $v$,
if $z<v<y_{ik-j}$ then
$\beta \in \mu_{i,j}(v)= \mu(v)$.

Possibly there are some $i'>0$ and $j'\in J$ such that
$0 \leq i'k-j' < ik-j$
( so
$y_{ik-j} \leq y_{i'k-j'+1} <
y_{i'k-j'} \leq y_0$)
and either one of the following
five holds:
$\beta \not \in \cover(b_{j'})$;
$\neg \alpha \not \in \cover(b_{j'})$;
$\beta \not \in \fin(b_{j'})$;
$\alpha \in \fin(b_{j'})$; or
$U( \alpha, \beta ) \not \in \fin(b_{j'})$.
If there is no such $i',j'$ then
it is straightforward to show that
R2.2 holds and we are done.
If there are such $i',j'$ then we can
suppose that they are chosen so that
$i'k-j'$ is greatest possible.
It follows that $U(\alpha,\beta) \in \start(b_{j'})$.
If R2.1 holds of $\mu_{i',j'}$ then 
it is easy to finish.
So suppose not.
Thus R2.2 holds of $\mu_{i',j'}$
and we can conclude via R6 that
$\beta \in \cover(b_{j'})$,
$\beta \in \fin(b_{j'})$
and $U(\alpha,\beta) \in \fin(b_{j'})$.
Because R2.1 does not hold of $\mu_{i',j'}$
we can also conclude via R6 that
$\neg \alpha \in \cover(b_{j'})$
and $\neg \alpha \in \fin(b_{j'})$.
This contradicts our choice of $i'$ and $j'$
and we are done.

For the converse direction of R2,
we assume that $z \in [x,y[$ and
either R2.1 or R2.2 holds.
The subcases concern whether
R2.1 or R2.2 holds and
whether
$z=x$, $z$ equals some $y_{ik-j}<y_0$
or $z$ is in some $]y_{ik-j+1},y_{ik-j}[$.
We must show that $U(\alpha,\beta) \in \mu(z)$.

Suppose R2.2
holds with $z$ in some $]y_{ik-j+1},y_{ik-j}[$.
So $\beta \in \mu(y)$,
$U( \alpha, \beta) \in \mu(y)$
and for all $v$,
if $z < v < y$ then
$\beta \in \mu(v)$.
A straightforward induction
on $i'k-j'$,
using R2.2 for each $\mu_{i',j'}$
shows that for all
such numbers with
$0 \leq i'k-j' \leq ik-j$,
we have
$U( \alpha, \beta) \in \mu(y_{i'k-j'})$.
That $U(\alpha,\beta) \in \mu_{i,j}(z)$
follows immediately
by using R2.2 on $\mu_{i,j}$.

For the forward direction of
condition R6,
suppose $\beta$ is in the cover of $m$.
Thus $\beta$ is in the start, end and
cover of each $b_i$ as they compose
(with $m$ itself) to $m$.
Also note that the end of $b_k$
is the same as the start of $b_1$.
By conditions R4, R5 and R6 for each $\mu_{i,j}$,
$\beta \in \mu(z)$ for each
$z \in [y_{ik-j+1},y_{ik-j}]$
as required.

For the converse direction
of condition R6,
suppose, for all $u \in ]x,y[$,
$\beta \in \mu(u)$.
It is clear that
$\beta$ is in the cover,
start and end of each
$b_i$.
If $\beta$ was not in the cover of $m$
then the fact that
$\langle m, b_1, ..., b_k \rangle$
is a full decomposition of $m$
would imply that
$\sim \beta$ would be in the start
of some $b_i$.
Hence, by contradiction,
$\beta$ is in the cover
of $m$ as required.
\end{proof}

Recall that a linear order $(T,<)$
is {\em separable}
iff there is countable set $Q \subseteq T$
such that if $s<t$ are from $T$ then
there is $q \in Q$ such that
$s<q<t$.
Clearly $\reals$ is separable
with $\ratn$ being a dense countable
suborder.

\begin{nemma}
{sepuffle}
Suppose $x<y$ are from $\unit$,
$0 \leq r$ and $0 \leq s$.

Then there are sets
$K_1, ..., K_r$ of closed intervals
of $]x,y[$ and 
sets $R_0, ..., R_s$
of elements of $]x,y[$
such that:
\begin{itemize}
\item if $[a,b] \in K_i$ and $[c,d] \in K_j$
and $[a,b]$ and $[c,d]$ are not disjoint
then $i=j$, $a=c$ and $b=d$;
\item if $[a,b] \in K_i$ then $[a,b]$ is disjoint from
$R_j$;
\item if $i \neq j$ then $R_i$ and $R_j$ are disjoint;
\item
if $u<v$ are from $[x,y]$ and are not both in the same
interval in some $K_i$ then for each $j=1,...,r$
there is an interval in $K_j$ which
begins strictly after $u$ and ends strictly before $v$
and for each $j=0,...,s$ there is some $z \in R_j$
such that $u<z<v$;
\item every $z \in ]x,y[$ appears in some $R_j$ or in
some interval in some $K_i$.
\end{itemize}
\end{nemma}

\begin{proof}
We can proceed in a two stage construction
as follows.
Stage one is the construction of the $K_i$.
If $r=0$ skip this stage.

Stage one proceeds in $\omega$ rounds
starting with round $0$.
Start with all the $K_i$ empty.
Before each round $K= \bigcup_{1 \leq i \leq r} K_i$ will
contain finitely many closed intervals
within $]x,y[$.
So there will be finitely many open 
maximal intervals partitioning the
complement of $\bigcup_{[u,v] \in K} [u,v]$
within $]x,y[$.
Call these the spaces left before that round.

In round 0 
put $[(2x+y)/3,(x+2y)/3]$ in $K_1$.
In general,
for each space $]u,v[$ left before 
round $pr+q+1$ (for integers $p\geq 0$ and $q$
with $0 \leq q <r$),
put $[(2u+v)/3,(u+2v)/3]$ in $K_{q+1}$.
Notice that we leave spaces
on each side of the new intervals
and these spaces are one third as
wide as the original space. 

After $\omega$ rounds we have our final
$K_i$s.

We will now use the 
separability property of $\reals$ to show
that there
are still plenty of points 
of $]x,y[$ not in any interval in any $K_i$.
Let $R$ be the set of these points.

I claim that between every pair of intervals
from $\bigcup K_i$ there are some elements of $R$.
To show this by contradiction suppose
that no element of $R$ lies between
$[a,b]$ and $[c,d]$ (where $b < c$).
So, for every $w \in ]b,c[$ there
is some interval
$I_{w} \in \bigcup K_i$
with $w \in I_w$.

Let $(S,\prec)$ be the ordering of intervals
from $\bigcup K_i$ 
which lie within $]b,c[$ 
inherited from their elements.
This is isomorphic to the rationals
order as it is 
countable, dense and without endpoints.

Thus the order $(S,\prec)$ has
an uncountable order $(G,\prec)$
of gaps.
Define a map $f:G \rightarrow \reals$
as follows:
given a gap $\gamma$ in $S$,
let
$X = \{ x \in \reals \mid x 
\mbox{ lies in some } [u,v] \in S \mbox{ with } 
[u,v] \prec \gamma \}$.
Let $f(\gamma) = \sup (X)$
which exists as $X \subseteq \reals$.
Clearly $f$ is order preserving
and one-to-one.
Furthermore,
if $\gamma < \delta$ are gaps of $S$
then there is $[u,v] \in S$
between them.
Thus
$b< f(\gamma) < u <v  < f(\delta) <c$
and $u$ must also be strictly between
the interval $I_{f(\gamma)}$ from $S$ containing $f(\gamma)$
 and the interval $I_{f(\delta)}$ from $S$ containing $f(\delta)$.
These two intervals must be disjoint.

Thus
$\{ I_{f(\gamma)} \mid \gamma \in G \}$
is an uncountable
set of pairwise disjoint
non-singleton intervals of $\reals$.
This clearly contradicts separability.

Thus $R$ is a set of points 
densely located between the 
intervals in the $K_i$
or, in the case that $r=0$,
$R=]x,y[$.

It is straightforward to partition
$R$ densely into the
pairwise disjoint $R_0, ..., R_s$ as required.

\end{proof}

\begin{nemma}
{shuffle}
If $m$ is fully decomposed by the tactic
$\shuffle( \langle P_0 , ..., P_s \rangle , 
\langle \lambda_1, ..., \lambda_r \rangle)$
with each mosaic in each $\lambda_i$
having a realization on any closed interval of $\unit$
then for any $x<y$ from $\unit$,
there is $\nu$ which realises
$m$ on $[x,y]$.
\end{nemma}

\begin{proof}
Let $K_1, ..., K_r$ 
and $R_0, ..., R_s$
be as constructed for $[x,y]$ in lemma~\ref{lem:sepuffle}.

Suppose $1 \leq i \leq r$ and
$\lambda_i = \langle s_1, ..., s_{e(i)} \rangle$.
For each interval $[u,v] \in K_i$
choose a sequence
$u= w_0 < w_1 < w_2 < ... < w_{e(i)}=v$
and for each $j=1, ..., e(i)$,
let $I(u,v,j)$ be the interval $[w_{j-1},w_j]$.
Let $\nu_{u,v,j}$ realize
$s_j$ on $I(u,v,j)$.

Define $\nu$ via
$\nu(x)=\start(m)$,
$\nu(y)=\fin(m)$,
for each $z$ in  $I(u,v,j)$
within an interval
$[u,v]$ from $K_i$,
$\nu(z)=\nu_{u,v,j}(z)$ and
for each $z \in R_i$,
$\nu(z)= P_i$.
Note that $z$ may lie at the end of some
$I(u,v,j)$ and the beginning of $I(u,v,j+1)$.
In that case, the fact that the
mosaics in each $\lambda_i$ compose
will guarantee that $\mu(z)$
is well-defined.

I claim that $\nu$ realizes $m=(A,B,C)$ on $[x,y]$.
Consider the six conditions.
The harder cases are conditions R2 and R6.
It is useful to consider condition R6 first.

For the forward direction of
condition R6,
suppose $\beta$ is in the cover of $m$.
By lemma~\ref{lem:shufcon} S0,
$\beta$ is in each $P_i$ and in the
start, cover and end of each mosaic in each $\lambda_i$.
So $\beta$ is in $\nu(z)$ for each 
$z$ in each $R_i$ and in each $I(u,v,j)$
in each $[u,v]$ in each $K_i$.
Thus $\beta \in \nu(z)$ for each
$z \in ]x,y[$
as required.

For the converse direction
of condition R6,
suppose, for all $u \in ]x,y[$,
$\beta \in \nu(u)$.
For contradiction suppose that
$\beta$ is not in the cover of $m$.
Thus $\sim \beta$ is a type 3 defect
in $m$ and this is cured in the
full decomposition F1.
Thus $\sim \beta$ appears in the start
of a mosaic in $\lambda$ or in $\mu$
or in the start of $m''$.
Thus $\sim \beta$ appears in the start
of a mosaic in one of the $\lambda_i$
or appears in one of the $A_i$, $C_i$
or $P_i$.
Thus $\sim \beta \in \nu(w)$ for
$w$ being the start of some $I(u,v,j)$
for some $[u,v]$ in some $K_i$
or for $w$ where some $[u,w]$ is in some $K_i$
or for $w$ in some $R_i$.
Thus $\beta$ can not be in $\nu(w)$
and we have our contradiction.
Hence, 
$\beta$ is in the cover
of $m$ as required.

To show the forward direction of R2
assume that $z \in [x,y[$ and $U(\alpha,\beta) \in \nu(z)$.
The subcases concern whether
$z=x$, $z$ is in some $R_i$
or $z$ is in some $I(u,v,j)$ for some $[u,v]$ in some $K_i$.
We must show that R2.1 or R2.2 holds.
Again there are several subcases
and their converses using similar sorts
of arguments.
We give a selection for illustration purposes.

First consider $z=x$. So
$U(\alpha,\beta) \in \nu(x) = A$.
Now $m$ is fully decomposed by
$\langle m' \rangle \cat \lambda \cat \mu \cat \langle m'' \rangle$
so, by definition of a full decomposition,
either
(1)
$U(\alpha,\beta)$ is a type 1 defect cured in the
decomposition or
(2) $\beta \in B$ and either
( $\beta \in C$ and $U(\alpha, \beta) \in C$)
or $\alpha \in C$.
These latter conditions (2)  give us the desired result
immediately.

If $U(\alpha,\beta) \in A$ is cured in the full decomposition
of $m$ then it is clear that $\beta$ is in the
cover of the first mosaic, $m'$.
But this cover is $B$ itself so $\beta \in B$
and $\beta \in \nu(v)$ for all $v \in ]x,y[$.
Now $\alpha$ appears in the end of a mosaic in the
full decomposition and so in $\nu(u)$
for some $u \in ]x,y[$.
Thus we are done.

Now consider the case of $z \in R_i$ with
$U(\alpha,\beta)\in \nu(z)=P_i$.
The case of $i=s$ is a slightly special case
of what follows and can be proved with
slight modifications so we will
omit that case. 
Assume $0 \leq i <s$.

Thus $y_i=(P_i,B,P_{i+1})$ is fully decomposed by
$\langle y_i, ..., y_s \rangle \cat \lambda \cat \langle y_0, ..., y_i \rangle$
and we must have $\beta \in \cover(y_i)=B$.
This is whether or not
$U(\alpha,\beta)\in P_i$ is a type 1 defect
in $y_i$ or not.
By the argument above for the R6 case,
$\beta \in \nu(v)$ for all $v \in ]x,y[$.

If $\sim \alpha \not \in B$
then $\alpha$ is a type 3 defect in $m$
and thus is cured in the
full decomposition.
Thus $\alpha$ appears in the start
of a mosaic in $\lambda$ or in $\mu$
or in the start of $m''$.
Thus $\alpha$ appears in the start
of a mosaic in one of the $\lambda_i$
or appears in one of the $A_i$, $C_i$
or $P_i$.
Thus $\alpha \in \nu(w)$ for
$w > z$ being the start of some $I(u,v,j)$
for some $[u,v]$ in some $K_i$
or for $w$ where some $[u,w]$ is in some $K_i$
or for $w$ in some $R_i$.
Combined with the observation about $\beta$
this gives us R2.1.

Otherwise, $\sim \alpha \in B$
and 
so coherency C1.2
along with the fact that $\neg U(\alpha, \beta) \not \in B$
gives us $U(\alpha, \beta) \in A$.
By the fullness of the decomposition
of $m$,
either 
$\alpha$ appears in the start
of a mosaic in $\lambda$ or in $\mu$
or in the start of $m''$
(and we proceed as above),
$\alpha \in C$ (and R2.1 holds)
or $\beta \in C$ and $U(\alpha,\beta) \in C$
(and R2.2 holds).
We are done.

The case of $z$
in some $I(u,v,j)$ for $[u,v] \in K_i$
is similar
but a little more complex.

For the converse direction of R2,
we assume that $z \in [x,y[$ and
either R2.1 or R2.2 holds.
The subcases concern whether
$z=x$, $z$ is in some $R_i$
or $z$ is in some $I(u,v,j)$ for some $[u,v]$ in some $K_i$.
We must show that $U(\alpha,\beta) \in \nu(z)$.

Consider just the case of R2.1 holding
for $z$ in some $I(u',v',j)$ for some $[u',v']$ in some $K_i$.
Let $u'=w_0 < w_1 < ... < w_{e(i)}=v'$ be such that
each $I(u',v',j)= [w_{j-1},w_j]$.
Thus $w_{j-1} \leq z \leq w_j$.
We have
$z< u \leq y$ and $\alpha \in \nu(u)$
and for all $v$,
$\beta \in \nu(v)$.
There are three possibilities for $u$.

Suppose $z < u \leq w_j$.
Thus R2.1 holds for $\nu_{u',v',j}$
and so $U(\alpha,\beta) \in \nu(z)=\nu_{u',v',j}(z)$.

Suppose $z \leq w_j \leq w_{j'} <  u \leq w_{j'+1} \leq w_{e(i)}$.
By R2.1 or R2.2, $U(\alpha,\beta) \in \nu(w_{j'})$.
An easy induction using R2.2 establishes that
$U(\alpha, \beta) \in \nu(w_j)$.
Then R2.2 gives us $U(\alpha,\beta) \in \nu(z)$ as required.

Suppose $w_{e(i)} < u \leq y$. 
For each $i=0,...,s$,
choose $w \in ]w_{e(i)},u[$ with
$w \in R_i$.
So $\beta \in \nu(w)=P_i$.
For each $i'=1,...,r$,
choose $[u'',v''] \in K_{i'}$
with $w_{e(i)} < u'' < v'' < u$.
Say $\lambda_{i'}=
\langle s_1, ..., s_{e(i')} \rangle$
and for each
$j'=1,...,e(i')$,
$I(u'',v'',j')=[w'_{j'-1},w'_{j'}]$.
Now 
$\beta \in \nu(w'_{j'-1})=\start(s_j)$,
$\beta \in \bigcap_{w'' \in [w'_{j'-1}, w'_{j'}]}
\nu( w'') = \cover(s_{j'})$ and
$\beta \in \nu(w'_{j'})=\fin(s_{j'})$.
We can conclude $\beta \in \nu(v')$
for all $v' \in ]x,y[$
and so by R6 that $\beta \in B=\cover(B)$
and so in the cover of all mosaics
in each $\lambda_{i'}$ and each
$x_{i'}$ and each $y_{i'}$.

There are two cases now:
either $\sim \alpha \in B$ or not.
Suppose $\sim \alpha \in B$ so that
$\sim \alpha \in \nu(w)$ for all
$w \in ]x,y[$.
So $\alpha \not \in \nu(w)$ for any such 
$w$. We know that $\alpha \in \nu(u)$
so it follows that $u=y$.
Thus $\alpha \in \fin(m)$.
Coherency C1.1 implies
that
$U( \alpha, \beta) \in \start(m)$
and  C2.2 gives
us $U(\alpha, \beta) \in \cover(m)=B$.
By R6, $U(\alpha, \beta) \in \nu(z)$ as required.

The other case is that
$\sim \alpha \not \in B$
so that $\alpha$ is a type 3 defect in $m$
and so appears
in the start of a mosaic in
some $\lambda_{i'}$, at the end
of some $\lambda_{i'}$ or
in some $P_{i'}$.
If $i<r$ then $x_i=(C_i,B,A_{i+1})$ is fully decomposed
by
$\langle x_i \rangle \cat 
\lambda_{i+1} \cat \langle x_{i+1} \rangle \cat
... \cat \lambda_r \cat \break
\langle x_r \rangle \cat
\mu \cat \langle x_0 \rangle \cat
\lambda_1 \cat \langle x_1 \rangle \cat ...
\cat \lambda_i \cat \langle x_i \rangle$.
If $i=r$ then $x_i=(C_i,B,P_0)$ is fully decomposed
by
$\langle x_r \rangle \cat \mu \cat \lambda$.
In either case $\beta$ is in the
starts, covers and ends of
all the mosaics and
$\alpha$ is in the start of one
of the mosaics. A simple induction
using coherency C1.1
tells us that
$U(\alpha,\beta) \in C_i$.

Now consider
$\lambda_i=
\langle s_1, ..., s_{e(i)} \rangle$ say with
$U(\alpha,\beta ) \in \fin(s_{e(i)})\rangle$
and $\beta$ in all the starts,
ends and covers.
A simple induction
using coherency C1.1
gives us $U(\alpha,\beta)$ in the start
and end of each $s_{j'}$.
So we have
$\beta$ and $U(\alpha,\beta)$ in 
$\nu_{i,j}(w_j)=\nu(w_j)$.
If $z=w_j$ we are done.
Otherwise, $w_{j-1} \leq z < w < w_j$.
However,
$\nu_{i,j}$ realizes $s_j$ on
$[w_{j-1},w_j]$ and so
condition R2.2 gives us
$U(\alpha,\beta) \in \nu_{i,j}(z)=\nu(z)$
as required.
\end{proof}

\begin{nemma}{rmssat}
Suppose that $\phi \in \lang$,
$q$ is an atom not appearing in $\phi$ and $m$ is
a $(\phi,q)$-relativized $*_q^\phi(\phi)$-mosaic which appears
in a real mosaic system.
Then $m$ is fully $\unit$-satisfiable.
\end{nemma}

\begin{proof}
let $*=*_q^{\phi}$.
Given 
the real mosaic system $S$ of
$*(\phi)$-mosaics,
we can easily proceed 
by induction on $k$ 
to show that,
for any $x<y$ from $\unit$,
for any level $k$ member $m \in S$ 
there is $\mu$ which realises
$m$ on $[x,y]$.
Each step of the induction is just a use of one or two
of the preceding lemmas
~\ref{lem:realcompo}, \ref{lem:lead},
its mirror image and \ref{lem:shuffle}.

So we have $\mu$ which realizes $m$ on $\unit$.
Define $h$ by $t \in h(p)$ iff $p \in \mu(t)$
and let ${\cal T}=(\unit,<,h)$.

I claim, for all $\alpha \in \clos (\sqpp)$,
for all $t \in \unit$,
\(
{\cal T},t \models \alpha
\mbox{ iff }
\alpha \in \mu(t).
\)

This is a straightforward
proof by induction on the construction of $\alpha$.
The case of $U(\alpha,\beta)$ is as follows.

Note that if
$U(\alpha,\beta) \in \clos (\sqpp)$
then $U(\alpha, \beta ) \leq \sqpp$
and so both $\alpha$ and $\beta$
are also in $\clos(\sqpp)$
by lemma~\ref{lem:inclosstar}.

First suppose
${\cal T},t \models U(\alpha,\beta)$.
Thus 
there is $s>t$ with
$0 \leq t < s \leq 1$,
${\cal T},s \models \alpha$ and
for all $u$, if
$t < u < s$ then
${\cal T},u \models \beta$.
By the inductive hypothesis,
$\alpha \in \mu(s)$ and
for all $u$, if
$t < u < s$ then
$\beta \in \mu(u)$.
By R2, $U(\alpha,\beta) \in \mu(t)$
as required.

Now suppose
$U(\alpha,\beta) \in \mu(t)$.
Note that $t< 1$ as
no $U(\gamma,\delta)$ is in $\mu(1)=\fin(m)$
as $m$ is relativized.
By R2, either
R2.1 or R2.2 holds
and it can not be the latter
as that entails $U(\alpha,\beta) \in \mu(1)$
amongst other things.
So R2.1 holds
and there is $s$ with
$t < s \leq 1$,
$\alpha \in \mu(s)$ and for all $u$,
if $t<u<s$ then
$\beta \in \mu(u)$.
It follows
via the inductive hypothesis that
${\cal T}, t \models U(\alpha,\beta)$ as required.

From the claim
and conditions R4, R5 and R6
on realization, it follows 
that
$\start(m)= \{ \alpha \in \clos(\sqpp)  | {\cal T},0 \models \alpha\}$,
$\fin(m)= \{ \alpha \in \clos(\sqpp)  | {\cal T},1 \models \alpha\}$
and
the cover of $m$ contains exactly those
$\alpha \in \clos(\sqpp)$ which
hold at all points in between $0$ and $1$.
Thus
$m = \mos_{\cal T}^\phi(0,1)$
as required.
\end{proof}

\section{Decomposition trees}
\label{sec:decomp}

In this section we begin to show
the converse of the last lemma,
to show that satisfiable mosaics
appear in real mosaic systems.
Here we show how to arrange
decompositions for satisfiable mosaics
into a tree structure.

\begin{nemma}
{realcomp}
Suppose ${\cal T}=([0,1],<,h)$ is a structure,
$\phi \in \lang$,
and
$0 \leq x_0 < x_1 < ... < x_n \leq 1$.

Then
the composition of
\(
\langle \mos(x_0,x_1), \mos(x_1,x_2), ...,
\mos(x_{n-1},x_n) \rangle
\)
is
$\mos(x_0,x_n)$.
\end{nemma}

\sfproof

\begin{nemma}
{satfd}
Suppose $\phi \in \lang$ and
${\cal T}=([0,1],<,h)$.
If $m=\mos(x,y)$ for some $x<y$ from 
$[0,1]$ then there is some sequence
$x=x_0 < x_1 < ... < x_{n-1} < x_n =y$
such that
$\langle \mos(x_0,x_1), ..., \mos(x_{n-1},x_n) \rangle$
is a full decomposition of $m$.
Furthermore,
the $x_i$ can be chosen
so that no $x_{j+1}-x_j$ is greater than
half of $y-x$.
\end{nemma}

\begin{proof}
We will choose a finite set of points
from $]x,y[$
at which we will decompose
$\mos(x,y)$.
For each defect $\delta$ in $\mos(x,y)=(A,B,C)$
choose some
$u_\delta$ or $z_\delta$ witnessing its cure
between
$x$ and $y$ as follows.

If $\delta = U(\alpha, \beta) \in A$ is
a type 1 defect then it is clear that
there must be
$u_\delta \in ]x,y[$ with
${\cal T}, u_\delta \models \alpha$ and
for all $v \in ]x,u_\delta[$,
${\cal T}, v \models \beta$.
Similarly find $u_\delta \in ]x,y[$
witnessing a cure for
type 2 defects.

If $\delta \in \clos \phi$ is a type 3 defect
in $\mos(x,y)$ then it is clear that
there is $z_\delta \in ]x,y[$ with
${\cal T}, z_\delta \models \delta$.

Collect all the $u_\delta$s and
$z_\delta$s so defined into a finite set
and add 
 the midpoint $(x+y)/2$
of $x$ and $y$.
Order these points between $x$ and $y$ as
$x=x_0 < x_1 < x_2 < ... < x_n < x_{n+1}=y$.
Note that some points might be in this
list for two or more reasons.

It is clear that
because of our choice of witnesses,
the 
sequence of $\mos(x_{j-1},x_j)$ is a full decomposition.
\end{proof}

\begin{definition}
$\;$\\

1. A {\em tree} here is
just a set (of nodes), partially-ordered
by a binary
irreflexive ancestor relation such that
the set of ancestors of any node
is finite and well-ordered (by the ancestor relation) and there
is a (unique) root (ie, ancestor
of every other node).

2. The {\em depth} of a node with $n$ ancestors
is $n+1$. 
%The {\em depth} of the tree is the maximum depth
%of any node in the tree.
So the root has depth $1$.

3. An {\em ordered tree} is a
tree with
finite numbers of children for
each node
and an earlier-later
relation which totally orders
siblings.

4. A {\em decomposition tree}
is an ordered tree with each node labelled
by  a pair $(x,y)$ of elements of $[0,1]$
such that $x<y$ and such that if node $g$ is labelled
by $(x,y)$ and has children labelled by
$(x_0,y_0), ..., (x_n,y_n)$ in order
then $x=x_0 < y_0 = x_1 < y_1=x_2 < ... < y_{n-1}=x_n < y_n=y$. 

5. A  decomposition tree
is {\em tapering} iff for all nodes $g$,
and children $h$ of $g$,
if $g$ is labelled by $(x,y)$
and $h$ is labelled by $(u,v)$ then
$v-u$ is at most half of $y-x$.

6. In the context of a structure $([0,1],<,h)$ and a
formula $\phi$, we say that
the $\phi$-mosaic $m$ is the
{\em mosaic label} of a node $g$
of a decomposition tree 
iff $g$ is labelled by $(x,y)$ and
$m = \mos(x,y)$.

7. If ${\cal T}=([0,1],<,h)$ is
a structure and
$\phi \in \lang$, then we say that
a decomposition tree 
is {\em $({\cal T},\phi)$-full}
iff for each node $g$,
if $g$ has children then
the mosaic labels on the children in order
form a full decomposition of the
mosaic label of $g$.

8. A $({\cal T},\phi)$-full decomposition
tree is {\em complete}
iff every node has children.

\end{definition}

In diagrams, we will represent earlier-later
by left-to-right ordering and
ancestors above descendents.

An  ordered tree has a depth-first earlier-later
total ordering of its nodes.
We will call this the {\em lexical} ordering
and sometimes restrict it
to leaf nodes.

\begin{nemma}
{satftdt}
Suppose ${\cal T}=([0,1],<,h)$ is a structure,
$\phi \in \lang$, and
$0 \leq x < y \leq 1$.
Then $\mos(x,y)$ is 
the mosaic label of the root of
a complete and tapering $({\cal T},
\phi)$-full decomposition tree. 
\end{nemma}

\begin{proof}
Say $m= \mos(t^-,t^+)$ for $t^-<t^+$ from $[0,1]$.
Construct a decomposition tree
with root labelled by $(t^-,t^+)$ by repeated use of
lemma~\ref{lem:satfd}.
\end{proof}

Consider the sequence of labels $(u,v)$
along any infinite branch $\eta$ of 
a tapering $({\cal T},\phi)$-full
decomposition tree $D$.
Because we have included the mid-points
of each $(u,v)$ in the labels
of the children of that node,
this sequence of pairs will converge
to some $r \in [0,1]$,
ie if the labels of nodes in order along $\eta$ are
$(x_0,y_0),(x_1,y_1),...$
then both sequences $x_0 \leq x_1 \leq ...$ and
$y_0 \geq y_1 \geq ...$ converge to $r$.
To see this, just note that the spread $(v-u)$
of a node's label is at most
half that of its parents. 
Call $r = \limit(\eta)$ the {\em limit} of $\eta$.

Note that if node $f$ is a child of node $g$
then the cover of the mosaic label
of $f$ includes (as a subset) the
cover of the mosaic label of $g$.
This is a simple property of compositions of mosaics.
The cover of the mosaic label of the
child may be strictly bigger.
However, it may be equal.
We are interested in infinite branches
in such a $D$ along which the cover of the 
mosaic labels remains the same
forever.

\begin{definition}
Suppose $H \subseteq \clos \phi$.
Say that the infinite branch
$\eta \subset D$ is an infinite  $H$-{\em branch}
iff
there is a node $e \in \eta$
labelled with a mosaic of cover $H$
such that
every node $f \in \eta$ which
is a descendant of $e$
is labelled with a mosaic with cover $H$.
\end{definition}

\begin{definition}
We say that the infinite branch $\eta$
lies after the node $k$ in an 
decomposition tree
iff
$k$ does not lie on $\eta$
and
there are some nodes of $\eta$ 
which lie lexically after $k$.

We say that the infinite branch $\eta$
lies before the node $k$ in an 
decomposition tree
iff
all the nodes of $\eta$ lie before $k$.

We say that the branch $\eta$ lies before the
branch $\theta$ iff
some node of $\theta$ lies after $\eta$.
\end{definition}

Note the slight asymmetry in the definitions
here reflecting the choice that
descendents of a node will
be lexically ordered after the node,
rather than before.

\begin{nemma}
{existsfirst}
 Suppose a node lies on an infinite $H$-branch.
Then there is a lexically first such branch
and a lexically last one.
\end{nemma}

\begin{proof}
To find the first branch, start at the node and recursively
move to the first child of the current
node which lies on an infinite $H$-branch.
\end{proof}

\begin{nemma}
{nextinf}
Suppose $k$ is a node labelled by $(k_-,k_+)$
in a decomposition tree
with root labelled by
$(g_-,g_+)$.

Then there is a sequence 
$k_+ = x_0 < x_1 < ... < x_n$
(possibly with $n=0$ and $k_+=x_0=x_n$)
with each
$(x_i, x_{i+1})$ the label of a leaf node of the
tree and either:
\begin{itemize}
\item $x_n=g_+$ and there are no infinite branches
after $k$, or
\item
there is a node $e$ labelled by $(e_-,e_+)$ 
lying on an infinite branch
with $x_n=e_-$.
\end{itemize}

There is a mirror image.
\end{nemma}

\sfproof

\begin{nemma}
{firstbranch}
Suppose that $E$ is
a tapering
$({\cal T},\phi)$-full decomposition tree
such that every sibling of a leaf is itself a leaf.

Suppose the root node $g \in E$ 
is labelled by $(x,y)$
such that $\mos(x,y)$ has cover $H$.
Also suppose that $g$ lies 
on an infinite $H$-branch.
Suppose $\eta$ is the lexically first 
infinite $H$-branch on which $g$ lies
and $\limit(\eta)=r$.

Then $x \leq r$.

If $x<r$ then
there
is a sequence 
$x=x_0 < x_1 < x_2 < ... < x_n \leq r$
such that $n>0$ and 
each $(x_i,x_{i+1})$ 
is the label of the parent of leaf node in $E$.

Furthermore,
the sequence can be chosen such that
if $x_n <r$ then there is a sequence
$x_n= y_0 < y_1 < ... < y_m < r$ 
such that $m>0$ and:
\begin{itemize}
\item
each $(y_i,y_{i+1})$ 
is the label of a leaf node in $E$;
\item
$\mos(y_0,r)$ is fully decomposed
by $\langle 
\mos(y_0,y_1), ..., \mos(y_{m-1},y_m), \mos(y_m,r) \rangle$;
\item
and $\mos(y_{m},r) = \mos(y_0,r)$.
\end{itemize}

There is a mirror image result with
$\eta$ being the lexical last branch on which $g$ lies.
\end{nemma}

\begin{proof}
Suppose that the mosaic $b$
appears infinitely often as a mosaic
label along $\eta$ in the tree $E$.
Thus the cover of $b$ is $H$.

For each $\delta= S(\alpha,\beta) \in \clos \phi$
such that ${\cal T},r \models \delta$
choose $u_\delta < r$ such that
${\cal T}, u_{\delta} \models \alpha$
and for all $w \in 
 ]u_{\delta}, r[$, 
${\cal T}, u_{\delta} \models \beta$.
We say that $\beta \in \clos(\phi)$ is
constantly true for a while before $r$
iff
there is some $x'<r$ such that
if $x' < w < r$ then
${\cal T}, x_{\beta} \models \beta$.
For each $\beta \in \clos \phi$ such that
$\beta$ is true for a while before $r$
choose some $x_\beta <r$ such that
if $x_{\beta} < w < r$ then
${\cal T}, x_{\beta} \models \beta$.
Now choose any node $k_0 \in \eta$
strictly below $g$
and labelled by $(u',v')$ such that
$u'$ is strictly greater than each
$u_\delta$ and each $x_\beta$.
This can be done as $v'-u'$ halves
in each generation but always
$u' \leq r \leq v'$.

Now say that $b$ appears as
$\mos(k_-,k_+)$ for some node $k \in \eta$
below $k_0$
and
labelled by $(k_-,k_+)$.

Let $f_1, ..., f_n$ be the sequence
of parents of leaf nodes of $E$ before $k$ in order.
By the mirror image of lemma~\ref{lem:nextinf}, there is 
a sequence $x=x_0 < x_1 < ... < x_n=k_-$ such that
the label of each $f_i$ is $(x_{i-1},x_i)$.
Note that we apply lemma~\ref{lem:nextinf}
to the subtree of $E$ without the leaf nodes
of $E$. In the statement of the lemma
we required that all siblings of leaves
are themselves leaves
in order to guarantee that this subtree
is a decomposition tree.

If $k_-=r$ then we are done.
Assume $k_-<r$.

We say that a formula $\gamma$ is true (in $\cal T$)
arbitrarily recently before a point $z \in T$
iff for every $z'<z$ there is some $z'' \in T$
with $z'<z''<z$ and ${\cal T}, z'' \models \gamma$.
For each $\beta \in \clos \phi$
which is true arbitrarily soon before $r$,
choose some $s_\beta$ such that
$k_- < s_\beta < r$ and
${\cal T}, s_\beta \models \beta$.
Now find a node $k'\in \eta$ below $k$
and labelled by $(k'_-,k'_+)$ such that 
 $\mos(k'_-,k'_+)=b$ 
and $k'_-$ is greater than each $s_\beta$.

Let $g_1, ..., g_p$
be the sequence
of parents of leaf nodes of $E$ below $k$
and before $k'$ in order.
Let $h_1, ..., h_{m-1}$ be the sequence
of children in $E$ of the $g_i$ in order. 
By the mirror image 
of lemma~\ref{lem:nextinf}
(used in the subtree rooted at $k$), there is 
a sequence $k_-=y_0 < y_1 < ... < y_{m-1}=k'_-$ such that
the label of each $h_i$ is $(y_{i-1},y_i)$.

If $k'_-=r$ then 
let $f_{n+1}, ..., f_{n'}$
be the parents of leaf nodes of $E$
below $k$ and before $k'$ in order.
By lemma~\ref{lem:nextinf}
(applied to the subtree of $E$ 
rooted at $k$ and not including the leaf nodes from $E$),
there is a sequence
$k_-=x_n < x_{n+1} < ... < x_{n'}=k'_- =r$
such that each
$(x_i,x_{i+1})$ is the label
of $f_i$.
The long sequence
$x=x_0 < x_1 < ... < x_n < ... < x_{n'}=r$
is as required and we are done.

Now assume $k'_-<r$.
Let $b' = \mos(k_-,r)$.
I claim that $\mos(k'_-,r)=b'$ too.
The start of both is the start of $b$.
The end of both is just
$\{ \alpha \in \clos \phi | {\cal T},r \models \alpha \}$.
Finally the cover of both is just
the set of $\beta \in \clos \phi$
such that $\beta$ holds constantly for a while
before $r$.

We will now show that
$\sigma= \langle \mos(y_0,y_1), ..., \mos(y_{m-1},y_m), \mos(y_m,r) \rangle$
fully decomposes $b'$.
The composition is $b'= \mos(y_0,r)$ by lemma~\ref{lem:realcomp}.

Before we show that the decomposition is full
consider the nodes below $k$.
Say that the children of node $k$ are
$k_1, ..., k_q$ labelled
by $(u_0,u_1),...,(u_{q-1},u_q)$
respectively.
Thus 
$\langle \mos(u_0,u_1), ..., \mos(u_{q-1},u_q) \rangle$
is a full decomposition of $\mos(k_-,k_+)=b$.
Also 
$k_-=u_0<u_1<...< u_{q-1} <u_q=k_+$
and there is some $j$ such that
$u_{j-1} \leq r < u_j$ and $k_j$ lies on $\eta$.
Because $\eta$ is the lexically first infinite
branch on which $k$ lies
there are only a finite number of nodes (in $E$) below
any $k_i$ with $i=1,...,j-1$.
Also, we can start with the
sequence $k_1, ..., k_{j-1}$ and repeatedly 
replace a node from $E$
by the sequence of its children in order
and end up with a prefix sequence
of $h_1, ..., h_{m-1}$ and a corresponding
sequence 
$k_-=u_0=y_0 < y_1 < y_2 < ... < y_M=u_{j-1}$.
Note that $u_0 < u_1 < ... < u_{j-1}$
will be a subsequence of this.

Now let us return to 
consider defects in $b'$.

{\bf Type 1 defects:}
Suppose $U(\alpha,\beta)\in \start(b')
=\start(b)$ is a type 1 defect
of $b'$.

If it happens that $\beta \not \in H=\cover(b)$
then $U(\alpha,\beta)$ is also
a type 1 defect in $b$
and thus cured in the
full decomposition
$\langle \mos(u_0,u_1), ...,
\mos(u_{q-1},u_q) \rangle$.
Thus the cure of $U(\alpha,\beta)$ is
witnessed in this sequence.
Say $\alpha \in \fin(\mos(u_{i-1},u_i))$
and
$\beta \in
\bigcap_{l=1}^{i-1}
( \cover(\mos(u_{l-1},u_l))
\cap
\fin(\mos(u_{l-1},u_l)))
\cap 
\cover(\mos(u_{i-1},u_i))$.
As $k_j$ labelled by $(u_{j-1},u_j)$ lies on $\eta$,
$\cover(\mos(u_{j-1},u_j))=H$ does
not contain $\beta$.
Thus $i<j$.
However, we have seen that then $u_i$ appears
as one of the $y_{i'}$
and thus we can find a witness to the
cure of the type 1 defect $U(\alpha,\beta)$ 
of $b'$ in the decomposition $\sigma$ as required. 

Now assume $\beta \in H
= \cover(b) \subseteq \cover(b')$.
For $U(\alpha, \beta)$ to be a type
1 defect in $b'$ we thus must have
$\alpha \not \in \fin(b')$ and either
$\beta \not \in \fin(b')$
or $U(\alpha, \beta) \not \in \fin(b')$.
Since $U(\alpha, \beta)$
holds at $k_-$
we must have $\alpha$ true
somewhere between $k_-$ and $r$.
So $\alpha$ is a type 3 defect of $\mos(k_-,k_+)$
and so cured in
$\langle \mos(u_0,u_1), ..., \mos(u_{q-1},u_q)\rangle$.
Thus $\alpha$ must
appear in the end
of $\mos(u_{i-1},u_i)$ say.
It can not appear after $r$ as
$U(\alpha,\beta)$ is not true at $r$
so $i<j$.
As above this implies
$\alpha$ appears in the end of a mosaic
in $\mos(y_{i'},y_{i'+1})$ in $\sigma$.

{\bf Type 2 defects:}
No type 2 defects are possible in $b'$.
Suppose $S(\alpha,\beta) \in \fin(b')$.
It is not possible that $\beta \not \in \cover(b')$
as $u_{S(\alpha,\beta)} < k_-$.
It is not possible that $\alpha \not \in \start(b')$
and $\beta \not \in \start(b')$
as $u_{S(\alpha,\beta)} < k_-$.
It is not possible that $\alpha \not \in \start(b')$
and $S(\alpha,\beta) \not \in \start(b')$
as $u_{S(\alpha,\beta)} < k_-$.

{\bf Type 3 defects:}
Suppose $\beta \in \clos \phi$
but $\sim \beta \not \in \cover(b')$.
So $\sim \beta$ does not hold constantly for a while
before $r$
and so is true arbitrarily recently before $r$.
So
$y_0=k_- < s_\beta < y_{m-1}=k'_-<r$
and ${\cal T},s_{\beta} \models \beta$.
Say that 
$y_{j'-1} \leq s_\beta < y_{j'}$.

So $s_\beta$ is within the label of $h_{j'}$
whose parent is $g_q$ say.
Maybe $\beta$ is in the
start or end of $\mos(a_-,a_+)$
where $(a_-,a_+)$ is the label of $g_q$.
Otherwise $\sim \beta \not \in \cover(\mos(a_-,a_+))$
and so $\beta$ appears in the end of a mosaic
in the full decomposition
of $g_q$.
So $\beta$
is witnessed
in $\langle \mos(y_0,y_1), ..., \mos(y_{m-2},y_{m-1}) \rangle$
as required.
\end{proof}

\begin{definition}
Say that the infinite $B$-branch
$\eta$ in a tapering
$({\cal T}, \phi)$-full decomposition tree
is a $B$-{\em stick}
iff
there is a node $e \in \eta$ 
which lies on only one infinite $B$-branch.
\end{definition}

\section{Satisfiability implies existence}
 
In this section we do the
main work of the paper and show
that satisfiable mosaics appear
in real mosaic systems
with a certain bound on the depth.

\begin{nemma}
{satrms}
Suppose $\psi \in \lang$ has length $L$ and
that $\psi$-mosaic $m_0$
is  $[0,1]$-satisfiable.
Then there is a real mosaic system
of depth $2L$
containing $m_0$.
\end{nemma}

\begin{proof}
Say
${\cal T}= ([0,1], < ,h)$,
$0 \leq t_0^- < t_0^+ \leq 1$ and
$m_0 = \mos^\psi_{\cal T}(t^-_0, t^+_0)$.
Let $S$ be the set of all
$\mos(x,y)$ for $x<y$ from $[0,1]$.
Clearly $m_0 \in S$.
I claim that $S$ is a real mosaic system
of depth $2L$.

In fact, I show that
for all $m \in S$,
for all $c= 0, ..., 2L$,
if $m$ has cover containing 
at least $2L-c$ formulas 
then $m$ is a level $c$ member of $S$.
We proceed by induction
on $c$.
Suppose that we have 
shown this for every $c' \leq c$
and mosaic $m =(A,B,C) \in S$ has 
cover containing at least $2L-(c+1)$ formulas.
All full trees will be
$({\cal T},\psi)$-full trees.

\begin{naim}
{lightening}
There is a tapering 
full decomposition tree $E$
with root with mosaic label $m=(A,B,C)$ such that:\\
\begin{tabular}{ll}
1. & all siblings of leaves are leaves,\\
2. & each leaf and each parent of a leaf is labelled by 
a level $c^+$ 
member of $S$,\\
3. & if a node of $E$ lies on an infinite branch
then its mosaic label has cover $B$,\\ 
and&\\
4. &  $E$ has no $B$-sticks in it.
\end{tabular}
\end{naim}

\begin{proof}
Choose any $g_-<g_+$ such that
$m=\mos(g_-,g_+)$.
Use lemma~\ref{lem:satftdt}
to find
a complete and tapering
full
decomposition tree $D$
with root with label $(g_-,g_+)$.

Let $E_0$ be the sub-tree 
of $D$ containing only 
the nodes with mosaic label with
cover $B$ and all their children and grandchildren.
Thus, any leaf node in $E_0$
and any parent of a leaf node in $E_0$
will have cover strictly
including $B$
and, by the inductive hypothesis
will be a level $c$ member of $S$.
Also, any sibling of a leaf node of $E_0$ will also
be a leaf node of $E_0$.

Enumerate the $B$-sticks in $E_0$.
This can be done as for each stick $\xi$
we can choose some node $e_\xi$
which lies on it and
on no other infinite $B$-branches.

We can use a step by step
process of gradually constructing
$E$ from $E_0$.
Each step removes one stick
$\xi$ by only changing the subtree of $E_0$
rooted at $e_\xi$.
The step introduces no
other infinite $B$-branches.
So it suffices to just show how to so remove
one $B$-stick $\xi$ 
from $E_0$ to make a tree $E'$.

Choose any node $f \in \xi$
below $e_\xi$ (so  
$f$ lies on no other infinite $B$-branches
apart from $\xi$).
Say that $f_1, ..., f_a$ are the
children of $f$ in order
and $f_d$ lies on $\xi$.
Say that $f_d$ is labelled by $(k_-,k_+)$
and that the limit of
$\xi$ is $r$ (so $k_- \leq r \leq k_+$).

To make $E'$ we will just
replace $f_d$ from $E_0$ and all its descendents
by a sequence of new children of $f$ who will be parents
of leaf nodes in $E'$.
The new children of $f$  will lie later than
$f_1, ..., f_{d-1}$ 
and earlier than $f_{d+1}, ..., f_a$.
In fact, we may replace $f_d$ by
one sequence of children 
with labels partitioning the interval
$[k_-,r]$ and another later sequence
with labels partitioning the
interval $[r,k_+]$.
Such a change can be seen to be 
as required in effecting
a removal of $\xi$ without any other
infinite branches being introduced
or even affected.
Note that as the start of the mosaic label
of the first new child of $f$ will just be
the same as the end of $f_{d-1}$
(or the start of $f$ in case that $d=1$),
and similarly for the end of the last new 
mosaic label,
the children of $f$ in $E'$ will still
carry  a full decomposition of the
mosaic label of $f$.

If $k_-=r$ or $r=k_+$ then we do not add
any new children in the first
or second sequence respectively.
Note that there will be some new children
to add in one or other or both sequences
as we can not have $k_-=r=k_+$.
Here we just show how to
construct the first sequence of new children
with labels partitioning
$[k_-,r]$ in the case that $k_-<r$.
Constructing the later second sequence is
via a mirror image argument.

So suppose $k_-<r$.
Lemma~\ref{lem:firstbranch}
applied to the subtree $E'_0$  of $E_0$  
consisting of
$f_d$ and all its descendents in $E_0$
tells us we have two cases.

Possibly there is a sequence
$k_-=x_0 < x_1 < ... < x_n=r$
such that each $(x_i,x_{i+1})$
is the label of a parent,  $g_i$ say, of a leaf node
in $E'_0$.
In this case the earlier sequence
of new children of $f$ in $E'$ will
be new nodes  $e_1, ..., e_{n+1}$
with each $e_i$ labelled by $(x_{i-1},x_i)$.
We also give each $e_i$ leaf node children
with exactly the same labels
as the leaf node children
of $g_i$.
We are done.

In the other case there is a sequence
$k_-=x_0 < x_1 < ... < x_n < x_{n+1}=r$
such that each $(x_i,x_{i+1})$ with $i<n$
is the label of a parent $g_i$ of a leaf node
in $E'_0$
and a sequence
$x_n=y_0 < y_1 < ... < y_m < y_{m+1}=r$
such that each $(y_i,y_{i+1})$ with $i<m$
is the label of a leaf node in $E'_0$
and
$\mos(y_m,r)=\mos(y_0,r)$ is fully
decomposed by
$\langle
\mos(y_0,y_1), ..., \mos(y_m,r) \rangle$.
In this case the earlier sequence
of new children of $f$ in $E'$ will
be new nodes  $e_0, ..., e_{n}$
with each $e_i$ labelled by $(x_{i},x_{i+1})$.
For $i<n$, we give each $e_i$ leaf node children
with exactly the same labels
as the leaf node children
of $g_i$.
Thus for $i<n$ each $\mos(x_i,x_{i+1})$
is a level $c$ member of $S$.
For the node $e_n$ labelled by $(x_n,r)$
we give it $m+1$ children,
$e'_0, ..., e'_m$ in that order.
We label each $e'_j$ by $(y_j,y_{j+1})$.
Now $\mos(y_0,r)=\mos(y_m,r)$ is fully decomposed by tactic
$\trail(\langle
\mos(y_0,y_1), ..., \mos(y_{m-1},y_m)\rangle)$
and each of these mosaics are mosaic labels
of leaf nodes in $E'_0$
and so are level $c$ members of $S$.
Thus $\mos(y_0,r)= \mos(y_m,r)$,
the mosaic label of both
$e_n$ and $e'_{m}$ is a level
$c^+$ member of $S$.
Again we are done.
\end{proof} 

Construct such an $E$
with root $g$ labelled by $(g_-,g_+)$.
If $E$ has no infinite branches then
it is clear that the mosaic labels
on the leaf nodes taken in lexical order
form a decomposition
of $m$. They are all in $S$ and so we have our
required decomposition.
Thus $m \in S$ is a level $(c+1)^-$
member of $S$ and hence trivially
a level $(c+1)$ member of $S$
and we are done.

We can thus assume that $E$ has two or more
infinite branches: a lone one would be a stick.
Say that $\eta_\infty$ is the lexically first one
and
$\theta_\infty$ is the lexically last one.

Possibly $g_-=\limit(\eta_\infty)$.
If not, ie if $g_-<\limit(\eta_\infty)$,  then
we can use lemma~\ref{lem:firstbranch}
to find either a sequence of 
level $c^+$ members of $S$ which
compose to
$\mos(g_-,\limit(\eta_\infty))$
or sequences $\sigma_0$ and $\rho_0$ 
of level $c^+$ members of $S$
and
a mosaic 
$b_1$,
such that
$b_1$ is fully decomposed by $\trail(\rho_0)$
and $\sigma_0 \cat \langle b_1 \rangle$ composes
to $\mos(g_-,\limit(\eta))$.

It follows that $\mos(g_-,\limit(\eta_\infty))$
is the composition of
level $(c+1)^-$ members of $S$.
Similarly, with $\theta_\infty$, 
and $g_+$.
We are done when we show that 
$o=\mos(\limit(\eta_\infty), \limit(\theta_\infty))$
is fully decomposed by
a shuffle of level $(c+1)^-$ 
members of $S$.
Then it follows that $\mos(g_-,g_+)$ is
a level $(c+1)$ member of $S$
as required.
Note that if $\limit(\eta_\infty)=\limit(\theta_\infty)$
then we would already be done,
so we are assuming
$\limit(\eta_\infty)<\limit(\theta_\infty)$.

Let $K= \{ \beta \in \clos \psi |
\sim \beta \not \in B \}$,
the set of type 3 defects in $m$.

Let $h_1$ be the deepest node on both
$\eta_\infty$ and $\theta_\infty$. 
Let $h_2$ be any descendent
of $h_1$ on $\eta_\infty$ which has two
children which each lie on
an infinite $B$-branch.
Such a node exists as $\eta_\infty$ is not a stick.
Let $h_3$ be a child of $h_2$
which does not lie on $\eta_\infty$
but does lie on another 
infinite $B$-branch.
Clearly $h_3$ has mosaic label with
cover $B$ and $h_3$ is lexically after
every node on $\eta_\infty$.

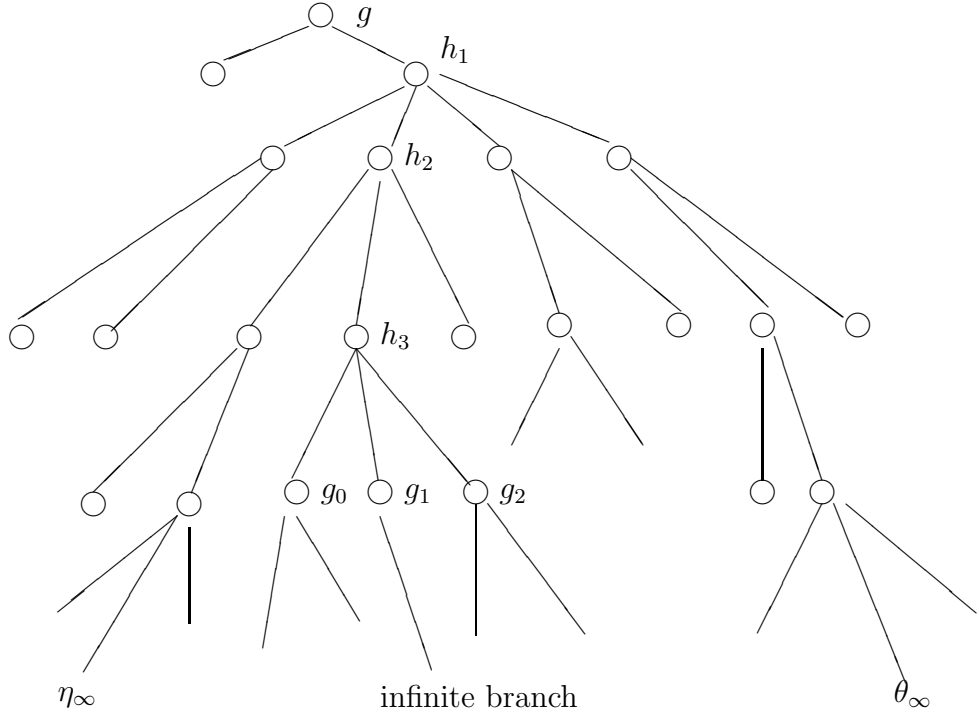
\begin{figure}
\begin{center}

\setlength{\unitlength}{3947sp}%
\begingroup\makeatletter\ifx\SetFigFont\undefined%
\gdef\SetFigFont#1#2#3#4#5{%
  \reset@font\fontsize{#1}{#2pt}%
  \fontfamily{#3}\fontseries{#4}\fontshape{#5}%
  \selectfont}%
\fi\endgroup%
\begin{picture}(6095,4487)(2393,-4240)
\thinlines
\put(4351,164){\circle{150}}
\put(3676,-211){\circle{150}}
\put(4951,-211){\circle{150}}
\put(4051,-736){\circle{150}}
\put(4726,-736){\circle{150}}
\put(5476,-736){\circle{150}}
\put(6226,-736){\circle{150}}
\put(2476,-1861){\circle{150}}
\put(3001,-1861){\circle{150}}
\put(3901,-1861){\circle{150}}
\put(4576,-1861){\circle{150}}
\put(5251,-1861){\circle{150}}
\put(5851,-1786){\circle{150}}
\put(6601,-1786){\circle{150}}
\put(7126,-1786){\circle{150}}
\put(7726,-1786){\circle{150}}
\put(4201,-2836){\circle{150}}
\put(4726,-2836){\circle{150}}
\put(5326,-2836){\circle{150}}
\put(2926,-2911){\circle{150}}
\put(3526,-2911){\circle{150}}
\put(7126,-2836){\circle{150}}
\put(7501,-2836){\circle{150}}
\put(4276, 89){\line(-5,-2){530.172}}
\put(4426, 89){\line( 2,-1){450}}
\put(4876,-286){\line(-2,-1){750}}
\put(4951,-286){\line(-2,-5){150}}
\put(5026,-286){\line( 6,-5){450}}
\put(5101,-211){\line( 5,-2){1060.345}}
\put(4651,-811){\line(-3,-4){738}}
\put(4726,-886){\line(-1,-6){150}}
\put(4801,-811){\line( 1,-2){480}}
\put(5551,-811){\line( 1,-3){300}}
\put(5551,-811){\line( 6,-5){1062.295}}
\put(6301,-811){\line( 1,-1){862.500}}
\put(6301,-736){\line( 4,-3){1332}}
\put(3976,-736){\line(-3,-2){1523.077}}
\put(4051,-811){\line(-1,-1){1012.500}}
\put(3826,-1936){\line(-1,-1){900}}
\put(3901,-1936){\line(-2,-5){362.069}}
\put(7126,-1936){\line( 0,-1){825}}
\put(7201,-1861){\line( 1,-3){300}}
\put(4576,-1936){\line(-1,-2){405}}
\put(4576,-1936){\line( 1,-6){137.838}}
\put(4576,-1936){\line( 5,-6){713.115}}
\put(3451,-2986){\line(-3,-5){588.971}}
\put(7576,-2911){\line( 2,-5){450}}
\multiput(4201,-2986)(61.08597,-101.80995){7}{\line( 3,-5){ 30.543}}
\put(3451,-2986){\line(-5,-4){750}}
\put(3526,-3061){\line( 0,-1){600}}
\put(4201,-2986){\line( 3,-5){397.059}}
\put(5401,-2911){\line( 3,-4){612}}
\put(5851,-1936){\line(-1,-2){300}}
\put(5926,-1861){\line( 2,-3){450}}
\put(7651,-2911){\line( 6,-5){818.852}}
\put(7501,-2911){\line(-1,-2){405}}
\put(4126,-2986){\line(-1,-6){137.838}}
\put(4726,-2986){\line( 1,-3){322.500}}
\put(5326,-2911){\line( 0,-1){825}}
\put(2701,-4186){\makebox(0,0)[lb]{\smash{\SetFigFont{12}{14.4}{\rmdefault}{\mddefault}{\updefault}$\eta_\infty$}}}
\put(7951,-4186){\makebox(0,0)[lb]{\smash{\SetFigFont{12}{14.4}{\rmdefault}{\mddefault}{\updefault}$\theta_\infty$}}}
\put(4876,-2911){\makebox(0,0)[lb]{\smash{\SetFigFont{12}{14.4}{\rmdefault}{\mddefault}{\updefault}$g_1$}}}
\put(4726,-4186){\makebox(0,0)[lb]{\smash{\SetFigFont{12}{14.4}{\rmdefault}{\mddefault}{\updefault}infinite branch}}}
\put(4576, 89){\makebox(0,0)[lb]{\smash{\SetFigFont{12}{14.4}{\rmdefault}{\mddefault}{\updefault}$g$}}}
\put(5101,-136){\makebox(0,0)[lb]{\smash{\SetFigFont{12}{14.4}{\rmdefault}{\mddefault}{\updefault}$h_1$}}}
\put(4876,-811){\makebox(0,0)[lb]{\smash{\SetFigFont{12}{14.4}{\rmdefault}{\mddefault}{\updefault}$h_2$}}}
\put(4726,-1936){\makebox(0,0)[lb]{\smash{\SetFigFont{12}{14.4}{\rmdefault}{\mddefault}{\updefault}$h_3$}}}
\put(4351,-2911){\makebox(0,0)[lb]{\smash{\SetFigFont{12}{14.4}{\rmdefault}{\mddefault}{\updefault}$g_0$}}}
\put(5476,-2911){\makebox(0,0)[lb]{\smash{\SetFigFont{12}{14.4}{\rmdefault}{\mddefault}{\updefault}$g_2$}}}
\end{picture}
\end{center}
\caption{A family of $g_i$s}
\end{figure}

Say that the children of $h_3$
are $g_0, ..., g_N$ in order.
Because the children
are labelled with a full
decomposition, each $\beta \in K$ appears
in the start of some $g_i$ for $i>0$.

\begin{naim}
{befaft}
For each $i=1, ..., N$, 
if $g_i$ is labelled by $(x,y)$,
 there are two infinite
$B$-branches $\theta'$ and $\eta'$ such that
$\limit(\theta') \leq x \leq \limit(\eta')$
and,
if $\limit(\theta') < \limit(\eta')$
then,
$\mos(\limit(\theta'),\limit(\eta'))$
can be decomposed
as a non-empty sequence of
level $(c+1)^-$ member of $S$
which includes
some mosaic with start or end 
equal to $\start(\mos(x,y))$.
\end{naim}

\begin{proof}
We find $\eta'$
and, if $x< \limit(\eta')$, a sequence $\mu$
of level $c^+$ members of $S$
which composes to $\mos(x, \limit(\eta'))$. 
Finding $\theta'$ 
and a similar sequence $\nu$ which composes
to $\mos(\limit(\theta'),x)$
is (almost)
a mirror image.
The sequence $\nu \cat \mu$ will be
as required.

Note that as $i \geq 1$,
$g_i$ will have a next earlier sibling
$g_{i-1}$
which will be labelled with $(w,x)$ for some $w$.

Use lemma~\ref{lem:nextinf}
applied to $g_{i-1}$
to find a node $e$ labelled
by $(e_-,e_+)$
lying on an infinite branch of $E$
and a sequence
$x=x_0 < ... < x_n=e_-$
with each $(x_i,x_{i+1})$
the label of a leaf node of $E$.
Note that there is an infinite branch
of $E$ which lies after $g_{i-1}$
as $\theta_\infty$ does.

Possible $x=e_-$ in which case
let $\sigma'$ be the empty sequence
of mosaics.
Otherwise, if $x<e_-$ then
let $\sigma'$ be the
sequence of $\mos(x_i,x_{i+1})$s
in order.
These are each level $c^+$ members of $S$
and the composition of the sequence
is $\mos(x,e_-)$.

Let $\eta'$ be the lexically first infinite branch on which $e$ lies.
If $e_-=\limit(\eta')$ then we are done
as $\mu=\sigma'$ will do.
So suppose that
$e_-< \limit(\eta')$.
We will find a sequence $\mu'$ of
level $(c+1)^-$ members of $S$
which compose to $\mos(e_-,\limit(\eta'))$
and then we can put
$\mu = \sigma' \cat \mu'$ and
we will be done.
By lemma~\ref{lem:firstbranch}
we have two cases.

Possibly there is a sequence
of mosaic labels of parents 
of leaf nodes in $E$ which
compose to $\mos(e_-,\limit(\eta'))$
and we can use that
as our $\mu'$.

The other possibility is that
we have a sequence $\tau$ of
mosaic labels of parents 
of leaf nodes in $E$ 
followed by one final mosaic $b$
such that
$\mu'= \tau \cat \langle b \rangle$
composes to
$\mos(e_-,\limit(\eta'))$
and
$b$ is fully decomposed by the
tactic $\trail(\langle \rho \rangle)$
where $\rho$
is a sequence of mosaic labels
of leaf nodes of $E$.
Again the mosaics in $\tau$
are level $c^+$ members of $S$
and $b$ is a level $(c+1)^-$ member
of $S$ and so we are done.
\end{proof}

Let $\{ (\theta_{-1},\eta_{-1}), ...,
(\theta_{-s},\eta_{-s})\}$
and $\{ (\theta_1,\eta_1),...,
(\theta_r,\eta_r) \}$
be the sets of all the pairs of infinite $B$-branches
got using claim~\ref{claim:befaft}
on each $g_i$ for $i=1,...,N$,
such that
each $\theta_{-j}$ and $\eta_{-j}$
have equal limits
but each $\theta_j$ and $\eta_j$
do not.
For each $j=1,...,s$,
let $P_j=\{ \beta \in \clos \psi |
{\cal T}, \limit(\eta_{-j}) \models \beta \}$.
For each $j=1,...,r$ let
$\lambda_j$ be a sequence
of level $(c+1)^-$ member of $S$
which compose to
$\mos(\limit(\theta_j),\limit(\eta_j))$.
By the  claim
and our original choice 
of $g_0, ..., g_N$, we can do this and ensure that
for each $\beta \in K$,
either there is a $P_j$ with
$\beta \in P_j$ or a $\lambda_j$
with $\beta$ in the start or end of some mosaic
in $\lambda_j$.

Now choose any infinite branch $\theta_0$
of $E$ as follows.
Start at $g$ and proceed recursively.
Choose a child of the current node
which lies on an infinite branch.
When there is a choice of such children
(as there will be infinitely often)
infinitely often choose an
earlier child,
infinitely often a later child.
Let $s$ be the limit of $\theta_0$.
Let $P_0 =
\{ \alpha \in \clos \psi |
{\cal T}, s \models \alpha \}$.
Also let $\eta_0=\theta_0$.

Say that an infinite $B$-branch $\kappa'$
is
{\em left dense} if 
for all nodes $e \in \kappa'$ there
is a descendent $f$ of $e$ in $\kappa'$
such that
$f$ has at least two children $f'$ and $f''$
on infinite $B$-branches such that
$f'$ is earlier than $f''$
but $f''$ lies on $\kappa'$.

Define {\em right dense} as the mirror image.

Note that due to the absence of sticks
each infinite $B$-branch is
either left dense or right dense or both.

\begin{naim}
{leftandright}
Each $\theta_j (-s\leq j\leq r)$
and $\theta_\infty$
is left dense
and each $\eta_j (-s \leq j \leq r)$
and $\eta_\infty$ is right dense.
\end{naim}

\begin{proof}
Each $\theta_i (i \neq 0)$ 
and $\theta_\infty$ is found
as the lexically last infinite $B$-branch
on which some node lies.
Thus it can not be right dense.
As it is not a stick it must
be left dense.
Similarly with $\eta_\infty$ and each $\eta_i (i \neq 0)$.

$\theta_0$ was chosen to be left dense and
right dense by construction.
\end{proof}

\begin{naim}
{atend}
If $U(\alpha, \beta)$
is true at the limit of a right dense
infinite branch of $E$
then $\beta$ is in the cover of $m$.
\end{naim}

\begin{proof}
Suppose $\eta'$ is a right dense infinite
branch with limit $s$ and
$P= \{ \gamma \in \clos \psi |
{\cal T}, s \models \gamma \}$.
If ${\cal T},s \models U(\alpha,\beta)$
then there is some $t>s$ such that
${\cal T},t \models \alpha$
and for all $u$,
if $s<u<t$ then
${\cal T},u \models \beta$.

Choose some node $h \in \eta'$
labelled by $(h_-,h_+)$
such that
$h_+ - h_- < t-s$
which we can do as the width of labels
halves with each generation.

Since $\eta'$ is right dense we
can choose some descendent $h'$ of $h$
with a child $f$ on an infinite $B$-branch
and an earlier child $f'$ lying on $\eta'$.
Say that $f$ is labelled with $(u,v)$.

Clearly $h_- \leq s \leq u < v \leq h_+ < t$.
Thus $\beta$ is in the cover of
$\mos(u,v)$ which is just $B$.
\end{proof}

\begin{naim}
{arby}
If $\alpha$ is a type 3 defect in $m$
and $\theta$ is a left dense infinite branch then
$\alpha$ is true arbitrarily recently before
$\limit(\theta)$.
There is a mirror image using right dense infinite branches
and arbitrarily soon afterwards.
\end{naim}

\begin{proof}
Let $\limit(\theta)=s$ and $t<s$.
Choose a node $n_1$ on $\theta$
labelled with
$(n_1^-,n_1^+)$
such that
$t < n_1^- \leq s$.
Choose a node $n_2$
on $\theta$ below $n_1$
with two children
$n_3$ before $n_4$
with $n_4$ on $\theta$ and
$n_3$ on another infinite branch.
So
$t < n_1^- \leq n_3^- < n_3^+ \leq n_4^- \leq s$.
Now the cover of $\mos(n_3^-,n_3^+)$ is $B$
and $\alpha$ is a type 3 defect
in $m$ and so in
$\mos(n_3^-,n_3^+)$.
Consider the full
decomposition
exhibited by the
children of $n_3$.
Thus there is some non-first child $n_5$ of
$n_3$ with
$\alpha \in \start( \mos(n_5^-,n_5^+))$.
Thus
$t < n_1^- \leq n_3^- < n_5^- < n_5^+ \leq n_3^+ \leq s$
and
${\cal T}, n_5^- \models \alpha$
as required.
\end{proof}

\begin{naim}
{coverob}
The cover of $o$ is $B$.
\end{naim}

\begin{proof}
As $g_- \leq \limit(\eta_\infty) < \limit(\theta_\infty) \leq g_+$,
the cover is contained in $B$.
For each $\beta \in \clos(\psi) \setminus B$,
$\sim \beta$ is a type 3 defect in $m$
and so  
by claim~\ref{claim:arby},
$\beta$ is true arbitrarily soon before
$\limit(\theta_\infty)$.
Thus $\sim \beta$ is also
not in the cover of $o$.
\end{proof}

\begin{naim}
{beforetheend}
If $\eta$
is a right dense
infinite branch of $E$
then $\limit(\eta) < \limit(\theta_\infty)$.
(And mirror image).
\end{naim}

\begin{proof}
It is clear that $\limit(\eta) \leq \limit(\theta_\infty)$.
We must rule out the case
of $\limit(\eta) = \limit(\theta_\infty)$.
Choose any node $n$ labelled by $(n_-,n_+)$ say on $\eta$ which has
a later sibling $n'$ labelled
by $(n'_-, n'_+)$ on another infinite branch $\kappa$ say.
Thus
$\limit(\eta) \leq n_+ \leq n'_- \leq \limit(\kappa')
\leq \limit(\theta_\infty)$.
Now choose a descendent $p$ of $n$ labelled by $(p_-,p_+)$
which lies on $\eta$ and has a later sibling
$p'$ lying on another infinite branch.
Thus
$\limit(\eta) \leq p_+ \leq p'_- < p'_+ \leq n_+ \leq \limit(\theta_\infty)$
as required.
\end{proof}

\begin{naim}
{rdfwd}
If $\eta$
is a right dense
infinite branch of $E$
then $Q= \{ \gamma \in \clos \psi |
{\cal T}, \limit(\eta) \models \gamma \}$ satisfies the
forward $K(o)$
property.
There is a mirror image.
\end{naim}

\begin{proof}
Suppose $\eta$ is a right dense infinite 
branch. First suppose
$U(\alpha,\beta)\in Q$.
By claim~\ref{claim:atend},
$\beta \in \cover(m)=B=\cover(o)$.

Now either $\alpha$ is a type 3 defect
in $m$
so $\sim \alpha \not \in \cover(m)$
(so K1)
or
$\alpha$ is not a type 3 defect
of $m$
so $\sim \alpha \in B$.
However,
$U(\alpha,\beta)$ is true at $\limit(\eta)$
so there is $s > \limit(\eta)$
with $\alpha$ true at $s$ and
$\beta$ true everywhere in
$]\limit(\eta),s[$.
Thus $s \geq g_+$
and $\beta$ is true everywhere
in $]g_-,s[$.
It follows that $\beta$ and $U(\alpha,\beta)$
hold everywhere in
$]g_-,s[$.
Now $g_- \leq \limit(\theta_\infty)
\leq g_+ \leq s$
so $\alpha$ holds at 
$\limit(\theta_\infty)$ if
$\limit(\theta_\infty)=g_+=s$
or
$\beta$ and $U(\alpha,\beta)$ hold
at $\limit(\theta_\infty)$
if $\limit(\theta_\infty) < s$.
Thus K2 or K3 holds as required.

To show the converse suppose that
$\beta \in B$ and K1, K2 or K3 holds
with respect to $U(\alpha,\beta)$ and the
forward $K(o)$ property.
Thus $\beta$ holds everywhere
between $\limit(\eta)$ and
$\limit(\theta_\infty)$.
If K2 or K3 holds then it is clear
that $U(\alpha,\beta)$
is true at $\limit(\eta)$.
If K1 holds then 
claims~\ref{claim:arby} and \ref{claim:beforetheend}
tell us that $\alpha$ is true
somewhere in between
$\limit(\eta)$ and
$\limit(\theta_\infty)$.
Again it follows that
$U(\alpha,\beta)$
is true at $\limit(\eta)$
as required.
\end{proof}

\begin{naim}
{finally}
The mosaic $o= \mos(\limit(\eta_\infty),
\limit(\theta_\infty) )$
is fully decomposed by
the tactic
shuffle
$( \langle P_0, ..., P_s \rangle,
\langle \lambda_{1},
...,
\lambda_{r}
\rangle )$.
\end{naim}

\begin{proof}
Let $A_i$ and $C_i$ be as in the definition of a shuffle.

We use lemma~\ref{lem:shufcon}.
All the necessary forward and backward $K(o)$ properties
(S1--S5) follow from claim~\ref{claim:rdfwd} above.

S0 holds by virtue of
claim~\ref{claim:beforetheend}.
S6 holds 
by choice of the $P_i$ and $\lambda_i$.
\end{proof}

This gives us our result
as 
$\mos(\limit(\eta_\infty), \limit(\theta_\infty))$ is 
fully decomposed by a
shuffle in which each mosaic in each $\lambda_i$
is a level $(c+1)^-$ 
member of $S$.
\end{proof}

\section{Summary so far}

Let us summarize.

\begin{definition}
Suppose $\psi \in \lang$.
Let $RMS(\psi)$ be the set of all
$\psi$-mosaics which
appear in any real mosaic system.
\end{definition}

\begin{nemma}
{levels}
Suppose $\psi \in \lang$.
Then $RMS(\psi)$ is
a real mosaic system
and the following are equivalent
for any $\psi$-mosaic $m$ and any
$n \geq 0$:\\
\begin{tabular}{ll}
1. & $m$ is a level $n$ member of
$RMS(\psi)$.\\
2. & $m$ is a level $n$ member of
some real mosaic system.\\
\end{tabular}
\end{nemma}

\begin{proof}
To show 2 implies 1 is straightforward
and it follows that
$RMS(\psi)$ is a real
mosaic system. It is then
clear that 1 implies 2.
\end{proof}

\begin{theorem}
\label{theorem:summary}
Suppose $\phi$ is a formula
of $\lang$ and  $q$ is an atom
not appearing in $\phi$.
Suppose $\psi = *^{\phi}_q(\phi)$
has length $N$.

Then the following are 
equivalent:\\
\begin{tabular}{ll}
1. & $\phi$ is $\reals$-satisfiable;\\
2. & there is a
$(\phi,q)$-relativized 
$\psi$-mosaic 
which appears in some real mosaic system;\\
3. & there is a
$(\phi,q)$-relativized 
$\psi$-mosaic 
which is a level $2N$ member
of $RMS(\psi)$.\\
\end{tabular}
\end{theorem}

\begin{proof}
{\bf (1 $\Rightarrow$ 3)}
If $\phi$ is satisfiable then
lemma~\ref{lem:satfmos}
implies there exists a
$(\phi,q)$-relativized $\psi$-mosaic
$m$ which is fully $\unit$-satisfiable,
and so is $[0,1]$-satisfiable.
Lemma~\ref{lem:satrms} implies $m$ appears in
a real mosaic system of depth $2N$.

{\bf (3 $\Rightarrow$ 2)} follows
from lemma~\ref{lem:levels}.

{\bf (2 $\Rightarrow$ 1)}.
If 
$(\phi,q)$-relativized $\psi$-mosaic
$m$ appears in a real mosaic system
then lemma~\ref{lem:rmssat} implies
that $m$ is fully $\unit$-satisfiable.
Thus lemma~\ref{lem:satfmos} tells
us that $\phi$ is $\reals$-satisfiable.
\end{proof}

\section{The width of the decompositions}
\label{sec:width}

In this section
we place bounds on the
number of mosaics needed in various
decompositions.
This is to allow
us to determine termination conditions
during nondeterministic algorithms.

Suppose $\phi \in \lang$ has length $L$.
There are at most $2L$ formulas in $\clos(\phi)$
and so there are at most
$2^{2L}.2^{2L}.2^{2L}=2^{6L}$
different $\phi$-mosaics.

\begin{nemma}
{shortcomps}
Suppose $\phi \in \lang$ has length $L$.
If the sequence $\sigma$ of $\phi$-mosaics
composes to $m$
then there is a subsequence $\sigma'$ of $\sigma$
of length at most $2^{7L+1}$
which also composes to $m$.
\end{nemma}

\begin{proof}
For each $\beta \in K
= \{ \beta \in \clos(\phi) |
\sim \beta \not \in \cover(m) \}$
choose a mosaic from $\sigma$
to witness $\beta$.
We can choose either
a non-first mosaic 
which has $\beta$ in its
start, 
a non-last mosaic which 
has $\beta$ in its end or 
any mosaic from $\sigma$
which does not have $\sim \beta$
in its cover.
Call these $\leq 2L$ mosaics the 
important ones in $\sigma$.
Construct $\sigma'$ by including
the important mosaics
and a composing subsequence of the
mosaics  between 
each consecutive pair of important mosaics
which contains no repeated mosaics.
A simple iterative procedure
allows us to successively remove
one copy of each repeat
and the mosaics in between.
Thus there will be at most $2^{6L}$
mosaics in $\sigma'$ in between
important mosaics.
The maximum length of $\sigma'$ will
be $2L. 2^{6L} \leq 2^{7L+1}$.
It is straightforward to check 
that the composition
of $\sigma'$ is $m$:
the cover is right because of the
inclusion of the important mosaics.
\end{proof}

\begin{nemma}
{shortlead}
If a $\phi$-mosaic $m$
is fully decomposed by the
tactic $\lead(\sigma)$
(or $\trail(\sigma)$)
then there is a subsequence $\sigma'$ of $\sigma$
of length at most $2^{7N+1}$
such that  $m$
is fully decomposed by the
tactic $\lead(\sigma')$
(or $\trail(\sigma')$ respectively).
\end{nemma}

\begin{proof}
Use the idea of important mosaics
as in the proof of the previous lemma
but include, as important, a witness
for the cure of each defect in $m$.
\end{proof}

\begin{nemma}
{shortshuffle}
If a $\phi$-mosaic $m$
is fully decomposed by the
tactic
shuffle
$( \langle P_0, ..., P_s \rangle,
\langle \lambda_{1},
...,
\lambda_{r}
\rangle )$
then 
$m$
is fully decomposed by a
tactic
shuffle
$( \langle P'_0, ..., P'_{s'} \rangle,
\langle \lambda'_{1},
...,
\lambda'_{r'}
\rangle )$
where
$r'+s' \leq 2L$,
each $P'_i$ is one of the $P_j$,
each $\lambda'_i$ is
a subsequence of one of the
$\lambda_j$
and each $\lambda'_i$
has length at most
$2^{7L+1}$.
\end{nemma}

\begin{proof}
By lemma~\ref{lem:shufcon}
we need only enough $P_i$
and $\lambda_i$
such that each
element of
$K= \{ \beta \in \clos(\phi) |
\sim \beta \not \in \cover(m) \}$
appears in some $P_i$
or in the start or end of a mosaic
in some $\lambda_i$.
Thus $r'$ and $s'$ can be chosen so that
$r'+s' \leq 2L$.

As in the proofs of the
previous lemmas
we can reduce each of the
chosen $\lambda_i$
to be of length $\leq 2^{7L+1}$
by removing repeats in between
important mosaics:
in this case
just the witnesses of 
elements of $K$.
\end{proof}

\section{RTL-SAT in PSPACE}
\label{sec:pspace}

Recall that we have
defined  RTL-SAT to be the problem
of deciding satisfiability
of formulas in the language
$\lang$
over real flows of time.
So, the idea is that we enter
a formula as input into
a machine and we get
a yes or no answer as output
corresponding to
satisfiability
or unsatisfiability respectively.
Here we show that RTL-SAT is in 
PSPACE.

We need to  specify 
how formulas of $L(U,S)$
are fed into a Turing machine.
There is a particular question
about the symbolic representation
of atomic propositions
since we allow them to be chosen from
an infinite set of atoms.
A careful approach
(seen in a similar example in
\cite{HoU79})
is to suppose (by renaming) that the
propositions actually used in a particular formula
are $x_1, ..., x_n$ 
and to code $x_i$ as the symbol $x$
followed by $i$ written in
binary.
Of course this means that the
input to the machine might be
a little longer than the
length of the formula.
In fact a formula of length $n$ may
correspond to an input of length
about $n \log_2 n$.
However, for a PSPACE algorithm
the difference is not enough for us to need to carefully distinguish
between the length of the
formula and the length of the
input.

In the proof we shall make
use
of non-deterministic Turing machines.
We use the definition
of NPSPACE (as in \cite{vEB90})
which requires all possible computations of
such a machine to terminate on any
input after using space polynomial  
in the size of the input.

\begin{definition}
We consider several algorithms,
each of which
is given a formula $\phi$ of $\lang$,
a natural number $n$,
and a $\phi$-mosaic $m$ 
(or more correctly a triple $(A,B,C)$
where $A,B$ and $C$
are subsets of $\clos(\phi)$).  
We say that 
a possibly nondeterministic algorithm
 is a $\phi$-NPSPACE one
iff
there is some polynomial $p(L)$
such that on any input with
$\phi$ of length $\leq L$ and
$n \leq 2L$
the algorithm returns a yes or no answer
after using at most $p(L)$ tape spaces. 
\end{definition}

\begin{nemma}
{LV}
There are $\phi$-NPSPACE algorithms which do the following
for each $\phi$, $n$ and $m$:
\begin{itemize}

\item
$SH(\phi,n,m)$ decides whether
or not there exists $P_i$s and
$\lambda_j$s such that
$m$ can be fully decomposed by the
tactic shuffle
$(\langle P_0, ..., P_s \rangle,
\langle \lambda_1, ...., \lambda_r \rangle)$
where each mosaic in each $\lambda_i$
is a level $n^-$ 
member of $RMS(\phi)$.

\item
$LV(\phi,n,m)$ decides whether
or not
$m$ is a level $n$ member of $RMS(\phi)$;

\item
$LD(\phi,n,m)$ decides whether
or not 
there is some $\sigma$ such that
$m$ can be fully decomposed by tactic
$\lead(\sigma)$
with each mosaic in $\sigma$
being a level $n$ member of $RMS(\phi)$;

\item
$TR(\phi,n,m)$ decides whether
or not
there is some $\sigma$ such that
$m$ can be fully decomposed by tactic
$\trail(\sigma)$
with each mosaic in $\sigma$
being a level $n$ member of $RMS(\phi)$;

\item
$CP(\phi,n,m)$ decides whether
or not
$m$ is a level $n^+$
member of $RMS(\phi)$;

\item
$LD'(\phi,n,m)$ decides whether
or not 
there is some $\sigma$ such that
$m$ can be fully decomposed by tactic
$\lead(\sigma)$
with each mosaic in $\sigma$
being a level $n^+$ member of $RMS(\phi)$;

\item
$TR'(\phi,n,m)$ decides whether
or not
there is some $\sigma$ such that
$m$ can be fully decomposed by tactic
$\trail(\sigma)$
with each mosaic in $\sigma$
being a level $n^+$ member of $RMS(\phi)$;

\item
$CM(\phi,n,m)$ decides whether
or not
$m$ is a level $(n+1)^-$ 
member of $RMS(\phi)$;

\end{itemize}
\end{nemma}

\begin{proof}
{\bf 1. Description of algorithms}.
The algorithms are defined in terms of each other.
First consider SH.
Given $\phi$ of length $N$, $n$ and $m$,
first check whether 
$m$ is a mosaic and check that
its start satisfies the forward $K(m)$ property
and its end satisfies the backwards $K(m)$ property.
Return the answer ``no'' if any of these checks
or subsequent checks fail.
Also collect the set $DEF$ of type 3 defects in $m$
and guess $s \leq 2N$.

For each $i=0$ to $s$,
guess $P_i$ and check that it is an MPC 
containing the cover of $m$ and satisfying
the forwards and backwards $K(m)$ property
and remove any $\beta \in P_i$
from the set $DEF$.

Guess $r \in \{0,1,..., 2N-s\}$.
For each $i=1,...,r$ (if any),
guess the start of the first mosaic in $\lambda_i$
and check that it satisfies the backwards $K(m)$ property
and guess the end of the last mosaic in $\lambda_i$
and check that it satisfies the forwards $K(m)$ property.
Also guess and check ``on the fly" a composing sequence $\lambda_i$
of up to $2^{6N+1}$ mosaics (with appropriate start and ends).
Check (via $CM(\phi,n-1,m')$)
that each $m'$ of these is a level $n^-$
member of $RMS(\phi)$
and remove from $DEF$ any formula
which appears in the start or end of $m'$.
Check that the start, cover and end of each $m'$
contains the cover of $m$.

Return ``yes" if $DEF$ ends up empty.
Otherwise return ``no".

Now consider LV.
To decide whether or not
$m$ is a level $n$ member of $RMS(\phi)$
we need to
guess a sequence of mosaics which compose
to $m$ and check that
each of these, $m'$ say,
is either a level $n^-$ 
member of $RMS(\phi)$
(so use $LV(\phi,n-1,m')$)
or
is fully decomposed by
a shuffle with
each mosaic in each sequence
in the shuffle
being a level $n^-$ member of $RMS(\phi)$
(so use $SH(\phi,n,m')$).

LD is as follows.
To decide whether
or not 
there is some $\sigma$ such that
$m$ can be fully decomposed by tactic
$\lead(\sigma)$
with each mosaic in $\sigma$
being a level $n$ member of $RMS(\phi)$,
we need to guess and check
a sequence $\sigma \cat \langle m \rangle$
which is a full decomposition of $m$
and check that each mosaic in $\sigma$
returns yes from $LV(\phi,n,m)$.

TR is similar to LD.

CP is  easy: we already know how to guess
and check decompositions.
LD' and TR'
are very similar to LD and TR.
CM uses 
LD' and TR' in the same
way CP uses LD and TR.
We already know how to guess
and check decompositions.

{\bf 2. The algorithms use 
polynomial space
and are correct.}
Fix $\phi$ of length $N$.
We proceed by induction on the number $n$ used.
Assume $n \geq 0$ and that we have shown that
the algorithms work for
any $n' < n$ and any $m$.

By lemmas~\ref{lem:shortshuffle} and \ref{lem:shufcon} 
and the inductive hypothesis,
$SH$ gives the correct result. 
By lemmas~\ref{lem:shortcomps} and \ref{lem:shortlead},
the other algorithms are correct.

The space bounds follow as each algorithm
needs only a small constant
amount of information about
each mosaic and the composition so far
in a possibly long composing sequence
of mosaics.
They may also need about $7N$ bits 
to represent, in binary,
the value of 
a counter as we check that 
the sequence is not too long.
Each call that they make to another algorithm
also requires a polynomial
amount of space but we know that
the depth of nesting of such calls
is just linear in $N$.
\end{proof}

We conclude
\begin{nemma}
{pspa}
RTL-SAT is in PSPACE.
\end{nemma}

\begin{proof}
An NPSPACE algorithm
is as follows.
Given $\phi$ of length $L$,
choose some atom $q$ not appearing
in $\phi$.
Guess a $(\phi,q)$-relativized 
$*_q^{\phi}(\phi)$-mosaic
$m=(A,B,C)$
(checking that it is is straightforward and uses
polynomial space).
Use LV from lemma~\ref{lem:LV}
to check whether
there is a real mosaic system
of depth $6L$
including  $m$.
By theorem~\ref{theorem:summary},
this approach gives ``yes'' answers
to satisfiable input
and the approach does not give incorrect ``yes'' answers.

By a theorem in \cite{Sav70} the problem is also
in PSPACE.
\end{proof}

\section{RTL-SAT is PSPACE-hard}
\label{sec:phard}

This part of the result is relatively straightforward.

\begin{lemma}
\label{lem:phard}
RTL-SAT is PSPACE-hard.
\end{lemma}

\begin{proof}
The proof of lemma 15 in \cite{Rey:cult}
(as one possible example amongst many
in the literature)
contains a formula
which we can easily modify.
The idea is to simulate the
running of any polynomial space bounded Turing machine
in a formula.

Let $M=(Q,\Sigma,\zeta,V_A,V_R,q_0)$
be a one-tape deterministic Turing Machine
 where
$Q$ is the set of states,
$\Sigma$ is the alphabet including blank $\blank$,
$\zeta : ( Q \times \Sigma ) \rightarrow
( Q \times \Sigma \times \{ L, R \})$,
$V_A \subseteq Q$ is the set of accepting states,
$V_R \subseteq Q$ is the set of rejecting states
and $q_0 \in Q$ is the initial state.
Suppose that $M$ is $S(n)$-space bounded,
where $S(n)$ is bounded by a polynomial in $n$.
Without loss of generality,
we may assume that $M$ is $2^{B(n)}$-time bounded 
where $B(n)$ is also bounded by a polynomial in $n$.
We also assume that  
once $M$ enters a state in $V_A$ (or $V_R$) then
it stays in states in $V_A$ ($V_R$ resp.).
Let $a= a_1 ... a_n$ be an input to $M$.

We can represent runs of $M$
via tape configurations in the usual way.
These may be supposed to be sequences
of $S(n)$ symbols each from $\Sigma \cup ( Q \times \Sigma )$.

We are going to effectively construct an $L(U)$ formula
$\phi$ which is of polynomial 
size in $n$
such that 
the satisfiability of $\phi$
is equivalent to
the acceptance of $a$ by $M$.

The atoms we use for $\phi$ are
from $Q \cup (Q \times \Sigma)
\cup \{ \tick, * \}$
along with 
$B(n)$ new atoms
$r_1, ..., r_{B(n)}$.

The idea of the proof
will be that $\phi$ is $\reals$-satisfiable
iff it is satisfiable 
in a certain  structure
$\cal T$
in this language.
$\cal T$ will represent an accepting run
of $M$ on $a$ in a straightforward way.
$\cal T$ has an initial tick point $0$ say. 
From then on, every tick point has a successor
tick point so we can name the points
$0,1,2,...$ etc but there may be
more tick points after those.
At every $(S(n)+1)$th tick point,
starting at $0$, 
the atom $*$ will hold.
The $S(n)$ tick points in between
$*$ points will represent
the contents of $M$'s tape
configuration at a particular instant.
The points $1,...,S(n)$ represent the
tape configuration
at the initial instant of $M$'s run
with input $a$.
For $1 < i \leq S(n)$,
the atom $a_i$ from $\Sigma$ will be true
at the $i$th point.
The atom $(q_0,a_1) \in Q \times \Sigma$
will hold at point 1.
The $S(n)$ points in between
the $*$ at point $S(n)+1$ 
and the $*$ at point
$2S(n)+2$ will similarly
contain the tape
configuration
at the second instant of $M$'s run.
And so on.

We will use the $r_i$s to count
up to $2^{B(n)}$ in binary at $*$ points
because we are only interested in the
first $2^{B(n)}$ steps in $M$'s computation.

The formula $\phi$ will be the conjunction
of $\phi_1, ..., \phi_{15}$ as defined below.
It should be clear that
$\phi$ is satisfiable iff
it is satisfiable in a model like $\cal T$
which represents a run of $M$ (on input $a$)
which is accepting,
iff $M$ accepts $a$.
That will complete our proof.

We use abbreviations
$\falsity= \neg \truth$,
$X \alpha= U( \tick \wedge \alpha, \neg \tick)$,
$F \alpha = U( \alpha, \truth)$ and
$G \alpha = \neg F ( \neg \alpha )$.
We also write
$X^{m+1} \alpha$ for $X X^m \alpha$
and $X^1 \alpha = X \alpha$.
Note that $F$ and $G$ are thus strict.

The discreteness of ticks is given by
$\phi_1=
\tick \wedge \neg S( \tick, \truth) \wedge
G ( \tick \rightarrow X \truth )$.
 
The distribution of $*$s is given by
$\phi_2= * \wedge X^{S(n)+1} *
\wedge G ( * \rightarrow
X^{S(n)+1} * )$.

$\phi_{3}$, which we will not write
out in detail
just prevents
any two different configuration symbol atoms
from
$Q \cup ( Q \times \Sigma ) \cup \{ * \}$
from holding at 
any one point and prevents
any of these symbols holding
at non-tick points.

The initial configuration is given by
$\phi_4 = \beta_0$ defined as follows.
Let $a_k= \blank$ for each $k>n$.
Now define each $\beta_k$ by recursion
down from $\beta_{S(n)}$ to $\beta_0$.
$\beta_{S(n)}=\truth$, each
$\beta_{k-1}= X( a_k \wedge \beta_k)$ $(k>0)$,
and $\beta_0 = X ( (q_0,a_1) \wedge \beta_1)$.
This is a formula of length
$< 5S(n)$.
 
The start of the second configuration is determined
by
$\phi_5= X^{S(n)+2} ( a' \wedge X (q',a_2))$
where $q'$ and $a'$ are such that
$\zeta(q_0,a_1) = (q',a',R)$.
(M must move right at first).

The relationship between a consecutive 
sequence of three symbols in 
any configuration
and the corresponding symbols 
at the next step is given in cases 
by $\phi_6, ..., \phi_{12}$.

$\phi_6 $
is the conjunction of all\\
\(
G (
( * \wedge X ( (q,a) \wedge X b ))
\rightarrow
X^{S(n)+1}
( * \wedge X ( a' \wedge X (q',b)))
)
\)\\
for each $q,q' \in Q,
a,b,a' \in \Sigma$
such that 
$\zeta(q,a)=(q',a',R)$.

$\phi_7$
is the conjunction of all\\
\(
G (
( a \wedge X ( (q,b) \wedge X c ))
\rightarrow
X^{S(n)+1}
( (q',a) \wedge X ( b' \wedge X c))
)
\)\\
for each $q,q' \in Q,
a,b,c,b' \in \Sigma$
such that 
$\zeta(q,b)=(q',b',L)$.

$\phi_8 $
is the conjunction of all\\
\(
G (
( a \wedge X ( (q,b) \wedge X c ))
\rightarrow
X^{S(n)+1}
( a \wedge X ( b' \wedge X (q',c)))
)
\)\\
for each $q,q' \in Q,
a,b,c,b' \in \Sigma$
such that 
$\zeta(q,b)=(q',b',R)$.

$\phi_9 $
is the conjunction of all\\
\(
G (
( a \wedge X ( (q,b) \wedge X * ))
\rightarrow
X^{S(n)+1}
( (q',a) \wedge X ( b' \wedge X *))
)
\)\\
for each $q,q' \in Q,
a,b,c,b' \in \Sigma$
such that 
$\zeta(q,b)=(q',b',L)$.

$\phi_{10} $
is the conjunction of all\\
\(
G (
( a \wedge X ( b \wedge X c ))
\rightarrow
X^{S(n)+1}
(  X  b )
)
\)\\
for each 
$a,b,c \in \Sigma$.

$\phi_{11}$
is the conjunction of all\\
\(
G (
( * \wedge X ( b \wedge X c ))
\rightarrow
X^{S(n)+1}
(  X  b )
)
\)\\
for each 
$b,c \in \Sigma$.

$\phi_{12}$
is the conjunction of all\\
\(
G (
( a \wedge X ( b \wedge X * ))
\rightarrow
X^{S(n)+1}
(  X  b )
)
\)\\
for each 
$a,b \in \Sigma$.

It is straightforward to show that
$\phi_5, ..., \phi_{12}$ along with
$\phi_{3}$
ensures
the progress of configurations
represented in any model of $\phi$
matches those of a run of $M$.

$\phi_{13}$
says that of the $r_i$s only
$r_1$ holds at time point
$S(n)+1$.

$\phi_{14}$
forces the $r_i$s to count
$*$ points.
This
large
conjunct of $\phi$
is still of size polynomial in $n$.
It is  
\[
G \bigwedge_{i=1}^{B(n)}
[ ( 
(* \wedge
\bigwedge_{j<i} r_j)
\wedge \neg r_i )
\rightarrow
(
X^{S(n)+1} (( \bigwedge_{j<i} \neg r_j ) \wedge r_i )
\wedge \bigwedge_{j>i}
( r_j \leftrightarrow
X^{S(n)+1}r_i) )].
\]

$\phi_{15}$
says that when the
$r_i$s are next all false
at an $*$ point then
from then on
the only $(q,a)$ atoms holding are those
with $q \in V_A$.
This forces the structure to be
representing
an accepting run of $M$
as described above.

\end{proof}

\section{Conclusion}
\label{sec:conc}

We have shown that the decision problem
for the temporal logic with  until
and since connectives
over real-numbers time
is PSPACE-complete.

There is a simple corollary 
using the expressive completeness
(\cite{GHR:book}) of RTL over the reals.
Consider a usual temporal logic,
ie 
one with connectives defined by first-order truth
tables as defined in \cite{GHR:book}.
It follows that
deciding any usual temporal logic
over the reals is a PSPACE problem
(but not necessarily PSPACE-hard).

In the introduction I suggested that
the mosaic proof here suggests a 
tableau style theorem-proving procedure
for the logic.
The idea would be to generate
all $\phi$-mosaics for a given $\phi$
and then systematically remove
those which can be decomposed
into simpler mosaics
(in the sense of a real mosaic system).
This would give an exponential time
procedure along the lines of that
seen in \cite{Pra79}.
We leave further development
of this idea as future work.

%\bibliography{../Templ}

\end{document}